\documentclass[format=acmsmall, review=false, screen=true,nonacm]{acmart}

\usepackage[english]{babel}
\usepackage{tabularx}
\usepackage{geometry}
\usepackage{amsmath}
\usepackage{graphicx}

\usepackage{hyperref}
\usepackage{enumitem}

\usepackage{array}
\usepackage{ragged2e}

\renewcommand{\arraystretch}{1.15}   
\usepackage{booktabs}

\usepackage{longtable}

\usepackage{multirow} 

\usepackage{float}

\title{From Prompting to Verification: How Experience Shapes Vibe Coding Practices}
\author{Ahmed Fawzy}
\email{a.fawzy@massey.ac.nz}
\affiliation{%
\
  \institution{School of Mathematical and Computational Sciences, Massey University}
  \city{Palmerston North}
  \country{New Zealand}
}

\author{Amjed Tahir}
\email{a.tahir@massey.ac.nz}
\affiliation{%
\
  \institution{School of Mathematical and Computational Sciences, Massey University}
  \country{New Zealand}
}

\author{Kelly Blincoe}
\email{k.blincoe@auckland.ac.nz}
\affiliation{%
  \institution{Department of Electrical, Computer, and Software Engineering, University of Auckland}
  \country{New Zealand}
}

\acmJournal{TOSEM}

\begin{abstract}
\AItools{} have expanded software creation beyond professional developers, giving rise to \emph{vibe coding}, a practice in which users generate software via natural-language prompts, evaluate outputs primarily by execution.
Prior work has examined how \AItools{} support programming tasks within specific user groups, typically professional developers, 
leaving open the question of how vibe coding practices differ across experience levels. 
We address this gap by surveying \AnalysedSampleSize{} vibe coders belonging to three user experience groups: non-coders, novices, and professional developers. 
Our results show that experience selectively shapes vibe coding. Reported experiences and perceptions of code quality are broadly similar across groups, with all three recognising both the strengths 
and limitations of vibe coding. In contrast, motivations, interaction styles, and quality assurance practices diverge with experience. Non-developers are most motivated by accessibility,  
novices emphasise learning and experimentation, and professionals use vibe coding more frequently in work-related contexts. 
We synthesise these findings as a \emph{perception--action gap}: a general awareness of risks in \aigc{} is broadly distributed, but the capacity to evaluate, debug, and verify remains experience-dependent. We show that vibe coding is partially democratising as it broadens access to software creation without equally distributing the expertise to evaluate it. 
\end{abstract}


\begin{document}

\newcommand{\fix}[1]{\textcolor{red}{#1}}
\newcommand{\add}[1]{\textcolor{blue}{#1}}
\newcommand{\Space}[1]{} 

\newcommand{\NumParticipates}{\fix{180}}
\newcommand{\NumParticipatesPilot}{\fix{9}}

\newcommand{\aigc}{AI-generated code}

\newcommand{\AItools}{AI code generation tools}
\newcommand{\vb}{vibe coding}

\newcommand{\RequiredSampleSize}{159}
\newcommand{\RecivedResponses}{181}
\newcommand{\AnalysedSampleSize}{162}

\newcommand{\writein}{30}

\newcommand{\pool}{665}
\newcommand{\poolnoneDev}{299}
\newcommand{\poolNovice}{228}
\newcommand{\poolProf}{138}

\newcommand{\poolRevok}{45}

\newcommand{\poolEng}{386}
\newcommand{\poolNotEng}{279}
\newcommand{\poolnoneDevEng}{105}
\newcommand{\poolNoviceEng}{163}
\newcommand{\poolProfEng}{118}

\newcommand{\poolEngInvitated}{260}

\newcommand{\matchedScreening}{96}
\newcommand{\retestSame}{71}
\newcommand{\retestSamePct}{74.0}
\newcommand{\retestChanged}{25}
\newcommand{\retestChangedPct}{26.0}
\newcommand{\retestToMiddle}{17}

\maketitle

\section{Introduction}
\label{intro}



\AItools{}, powered by large language models (LLMs), have rapidly become embedded in software development. Since the release of GitHub Copilot in 2021 and the public availability of conversational assistants such as ChatGPT in late 2022, these tools have evolved from inline autocomplete-style suggestions into interactive, multi-turn workflows in which coders describe functionality in natural language and iteratively refine \aigc{} through prompt--response cycles~\cite{barke2023grounded,imai2022github}. A further shift occurred in early 2025 with the emergence of CLI-based agentic coding tools such as Cursor, Claude Code, and OpenAI Codex, which can autonomously plan, execute, and iterate on multi-step development tasks with limited manual intervention~\cite{galster2026configuring,horikawa2025agentic}. Adoption has scaled accordingly. For example, the 2025 Stack Overflow Developer Survey reports that 84\% of respondents use or plan to use \AItools{} in their development process, up from 76\% in 2024, with 51\% of professional developers use them daily~\cite{stackoverflow2025}. Empirical studies report measurable productivity gains for routine programming tasks~\cite{peng2023impact,ziegler2024measuring}, although recent randomised evidence with experienced open-source developers finds that perceived productivity gains do not always match objective task completion times~\cite{becker2025measuring}. Persistent concerns also remain regarding code quality~\cite{yeticstiren2023evaluating}, security~\cite{sandoval2023lost,fu2025security}, overconfidence in outputs~\cite{perry2023users}, and mismatches between user expectations and tool outcomes~\cite{vaithilingam2022expectation}.

In parallel, the user base of \AItools{} has broadened far beyond professional developers. Non-developers, students, and individuals with little or no programming background can now produce functional code by describing goals in natural language~\cite{feldman2024non,prather2023s,kazemitabaar2023studying}, and practitioner accounts describe non-developers building applications for personal, organisational, and public-sector use~\cite{fedtec2025vibe,yang2025noncoders}. One industry report estimates that roughly a quarter of recent Y Combinator startups have codebases written almost entirely by AI~\cite{mehta2025yc}. This widening access creates a tension that prior work has begun to surface, but has not yet been examined across experience levels. End users without formal training often cannot reliably detect errors in \aigc{} even when explicitly warned to do so~\cite{virk2025nonprogrammers}, and tend to rely on whether the code runs as a proxy for correctness~\cite{vaithilingam2022expectation,perry2023users}. This widening user base has been accompanied by the emergence of new programming practices that differ from traditional AI-assisted programming.

One such practice is \emph{vibe coding}, a programming practice in which users generate software through natural-language prompting, evaluate outputs primarily through execution, and iterate via reprompting rather than systematic debugging~\cite{fawzy2025vibe}. Unlike conventional AI-assisted programming, where developers retain deliberate control over planning and verification ~\cite{barke2023grounded,mozannar2022reading}, vibe coding emphasises outcome-driven `trial-and-error' practice ~\cite{karpathy2025vibe,fawzy2025vibe}.
Vibe coding lowers barriers to software creation by enabling users, including those with little or no programming experience, to build software through natural-language interaction with \AItools{}, supporting rapid prototyping, learning, and creative exploration~\cite{fawzy2025vibe,kazemitabaar2023studying}. However, this accessibility comes with a recurring speed--quality trade-off, as code inspection and structured quality assurance are often reduced~\cite{fawzy2025vibe}. A key concern is how users respond to this trade-off. Although only 33\% of developers report trusting the accuracy of \aigc{}~\cite{stackoverflow2025}, reliance on these tools continues to grow, and user studies show that developers can articulate awareness of tool limitations without always translating this awareness into systematic verification practices~\cite{wang2024investigating,liang2024large}.
This creates a clear tension, as software becomes easier to produce but not necessarily to evaluate, debug, or secure.

Since different user groups engage in vibe coding practices, it is important to understand how different groups interact with it. Prior work has examined AI-assisted programming within specific user groups (i.e., professionals or students~\cite{liang2024large, vaithilingam2022expectation}), or studied non-experts in isolation~\cite{feldman2024non, prather2023s}. More broadly, AI code generation has often been studied as a tool or workflow-centred phenomenon, with emphasis on productivity~\cite{peng2023impact,ziegler2024measuring,becker2025measuring}, usability~\cite{vaithilingam2022expectation,liang2024large}, code quality~\cite{yeticstiren2023evaluating,liu2023your,nguyen2022empirical,tambon2025bugs}, and interaction patterns~\cite{barke2023grounded,mozannar2022reading,tang2024developer}, rather than as a coding practice shaped by users' programming backgrounds. As a result, it remains unclear how motivations, experiences, perceived code quality, and quality assurance (QA) practices differ across experience levels in vibe coding contexts. More importantly, it is also unclear whether recognising problems in \aigc{} leads to effective corrective action, or whether a gap exists between awareness and behaviour. This gap matters because if users share awareness of code quality issues but differ in how they respond to them, the risks of vibe coding depend not only on AI tools but also on users' capabilities. Characterising these differences can reveal where verification breaks down, identify which users are most vulnerable to relying on unverified \aigc{}, and inform the design of experience-aware tools, education in code evaluation and debugging, and organisational practices for safer adoption. To the best of our knowledge, no study has systematically compared vibe coding practices across non-developers, novice developers, and professional developers within a unified design.

To address this gap, we conduct a survey of those engaged in vibe coding practices, examining it as a cross-experience behavioural phenomenon. The study is grounded in themes identified by our prior grey literature review of practitioners' blogs, forums, and technical reports that characterised vibe coding along four behavioural dimensions: \emph{motivations} (why users engage in the practice), \emph{experiences} (what they encounter while doing so), \emph{perceived code quality} (how they judge the resulting code), and \emph{QA practices} (how they verify or validate it) \cite{fawzy2025vibe}. Our previous review offered only a qualitative account synthesised from practitioner discourse, finding vibe coding was driven largely by speed and accessibility and accompanied by informal QA, but it did not measure how these behaviours varied across users with different programming backgrounds. We address this by operationalising the four dimensions as quantitative, Likert-based constructs and examining how they differ across three user groups: non-developers, novice developers, and professional developers. We collected data from \RecivedResponses{} participants, of which \AnalysedSampleSize{} met the data quality criteria and were evenly distributed across the three groups (54 each).


Our results reveal a consistent pattern that experience shapes some aspects of vibe coding behaviour but not others. Participants across all experience levels report similar experiences and perceptions of \aigc{} quality, indicating a shared awareness of both the benefits and limitations of vibe coding. In contrast, clear differences emerge in motivations, interaction styles, and QA practices. Non-developers are most motivated by accessibility and empowerment and more often delegate code generation entirely to the AI; novices are more motivated by learning and experimentation; and professionals engage in more iterative dialogue, provide richer contextual input, and use vibe coding more often in work-related contexts. The clearest divergence is in QA, with less experienced participants relying more on reprompting rather than manual debugging and reporting greater difficulty understanding \aigc{} when errors arise, with non-developers the only group reporting that they \emph{never} check \aigc{} before use, whereas approximately 45\% of professionals report \emph{always} checking it.

These findings suggest a potential \emph{perception--action gap}. All user groups share awareness of the \aigc{} limitations, but their ability to respond to them varies with experience.
The risks of vibe coding are therefore shaped not only by the \AItools{}, but also by differences in user capability across experience levels. This has direct implications for how vibe coding should be supported. \AItools{} should provide stronger scaffolding for less experienced users (e.g., proactive error surfacing, explanatory support, and uncertainty signalling) while remaining unobtrusive for experts, and education should treat the evaluation of \aigc{} as a distinct skill alongside prompting.


\section{Background and Related Work}
\label{related-work}

\subsection{AI-Assisted Programming and Code Generation}
 
The rapid integration of \AItools{} represents a significant transformation in software development practices~\cite{chen2021evaluating,ziegler2024measuring}. AI code generation and conversational AI assistant tools have become embedded in development workflows, supporting code generation, completion, and debugging tasks~\cite{imai2022github, zhang2023practices, ziegler2022productivity}. These tools leverage LLMs trained on large-scale code corpora to generate contextually relevant code from natural-language prompts, enabling rapid, iterative development cycles~\cite{chen2021evaluating, fried2022incoder, wang2021codet5}. Recent empirical studies show that while \AItools{} improve productivity for routine programming tasks, their performance varies depending on task complexity and context~\cite{nguyen2022empirical, vaithilingam2022expectation}. Developers often use these tools to reduce keystroke effort, accelerate development, and recall syntax or APIs~\cite{liang2024large}. Interaction patterns extend beyond simple code completion; developers engage in iterative prompting and refinement workflows that shift attention from line-by-line coding to higher-level task specification~\cite{barke2023grounded}. However, even with these tools available, developers frequently refine or discard AI-generated suggestions. Recent studies have consistently report a gap between how useful developers expect AI tools to be and how effectively they integrate them into their actual workflows~\cite{vaithilingam2022expectation, liang2024large}, a pattern that persists even as \AItools{} have grown considerably more capable~\cite{becker2025measuring, chen2026beyond}. 
Notably, prior work has focused primarily on two groups: professional software developers and computer science students~\cite{barke2023grounded, liang2024large, nguyen2022empirical, vaithilingam2022expectation, ziegler2022productivity}.
It remains unclear whether the productivity benefits, interaction patterns, and failure modes extend to users who lack formal programming training (i.e., non-developers) -- a population that has grown substantially as \AItools{} have become widely accessible~\cite{feldman2024non, scaffidi2005estimating}. 

\subsection{Experience and Human--AI Interaction in Programming}
The integration of AI assistants into development workflows has introduced new patterns of human--AI interaction, shifting the developer's role from writing code directly to specifying, evaluating, and refining AI-generated outputs~\cite{barke2023grounded, mozannar2022reading}. This interaction model redistributes cognitive effort; rather than reasoning only about implementation, users increasingly need to assess and steer AI-generated code suggestions~\cite{liang2024large}. Observational research shows that this shift is also reflected in how developers direct their attention, with reviewers of \aigc{} tending to scan the high-level structure rather than closely examine implementation details~\cite{tang2024developer}.

This shift is particularly important because AI-assisted programming now involves users with varying levels of software development experience. Prior research on end-user programming has long shown that people without formal software engineering training create software artefacts to support their own goals~\cite{myers2006invited, nardi1993small, scaffidi2005estimating}. Recent studies suggest that \AItools{} extend this pattern by enabling non-expert users to produce functional code for practical tasks~\cite{feldman2024non, kazemitabaar2023studying}. However, these users may lack the conceptual knowledge needed to assess whether generated code is correct, robust, or maintainable~\cite{prather2023s}, and may rely on surface-level indicators, such as whether the code runs without error, as a proxy for correctness~\cite{vaithilingam2022expectation, perry2023users}. These concerns are especially relevant in AI-assisted contexts, where \aigc{} may appear correct while concealing deeper issues~\cite{denny2024computing, liang2024large}.

Experience level also shapes how these patterns play out in practice. Novice developers benefit from reduced syntactic burden and faster feedback cycles, but are less able to critically evaluate the outputs they receive~\cite{peng2023impact, prather2023s}. Professional developers, despite being more capable of identifying problems, still exhibit reliance on AI assistance even for tasks within their expertise~\cite{bird2022taking, ziegler2024measuring}. Across experience levels, studies reported that developers recognise potential risks in \aigc{} yet do not consistently adjust their validation behaviour in response~\cite{perry2023users, vaithilingam2022expectation, wang2024investigating}. However, these findings are drawn mainly from separate studies targeting distinct populations, typically either professional developers or computing students, rather than from designs that directly compare non-developers, novices, and professionals~\cite{liang2024large, prather2023s, vaithilingam2022expectation}. 
\subsection{Code Quality of AI-Generated Code}
 
The quality of \aigc{} is a persistent concern~\cite{yeticstiren2023evaluating}. While AI assistants can produce syntactically valid and functional code for straightforward tasks, empirical evaluations show that non-functional properties (such as security, maintainability, and correctness under edge cases) remain problematic~\cite{nguyen2022empirical, tambon2025bugs, liu2023your}. Recent longitudinal evidence reinforces this concern at the project level. A difference-in-differences study of open-source repositories adopting an agentic AI coding assistant found that short-term gains in development velocity were accompanied by a substantial and persistent rise in code complexity and static analysis warnings, with the accumulating technical debt in turn associated with slower development over time~\cite{he2026speed}. Security risks have received particular attention. Previous research shown that \aigc{} contain security weaknesses, sometimes in large numbers~\cite{pearce2025asleep, fu2025security}, and
users tend to rate \aigc{} as more secure than it actually is~\cite{perry2023users}. Novice users are less able to critically evaluate \aigc{} and more likely to accept suggestions without understanding them~\cite{prather2023s, denny2024computing}, suggesting that overconfidence about security may be particularly pronounced among less experienced users. Despite the practical importance of this possibility, no study has directly measured how the confidence--quality gap varies across user experience levels in an AI-assisted coding context. Quality assurance practices in AI-assisted development also differ from traditional workflows~\cite{vaithilingam2022expectation, liang2024large}. Rather than applying systematic testing or structured code review \cite{bacchelli2013expectations}, developers frequently rely on running the code to see whether it works, treating successful execution as sufficient validation~\cite{vaithilingam2022expectation}. Approaches such as test-driven development can help constrain the scope of \aigc{}~\cite{fakhoury2024llm}, but their adoption in AI-assisted workflows remains limited~\cite{fakhoury2024llm,mathews2024test}. 

\section{Methodology}
\label{Method}

\begin{figure}
    \centering
    \includegraphics[width=1\linewidth]{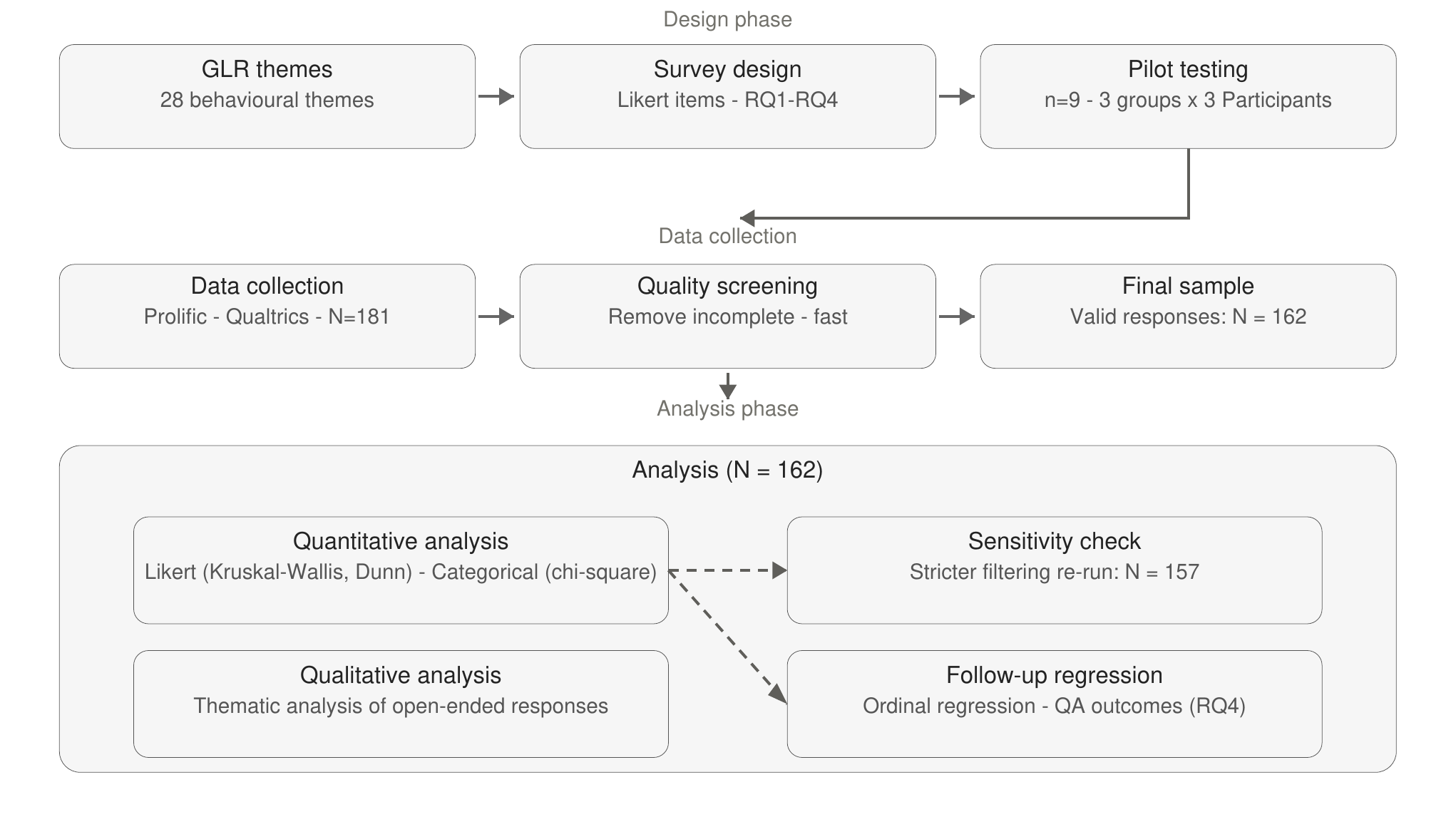}
   \caption{An overview of the questionnaire workflow}
\label{fig:study_workflow}
\end{figure}

In this study, we collected data from practitioners through an online questionnaire. to investigate how \emph{vibe coding} practices vary across coders with different levels of programming experience. The questionnaire was designed following established guidelines for survey research in software engineering. We drew on the principles for conducting online questionnaires in software engineering described by Punter et al.~\cite{punter2003conducting} and the structured questionnaire process proposed by Lin{\aa}ker et al.~\cite{linaaker2015guidelines}. These guidelines informed the selection practices relevant to our study (e.g., survey design, pilot testing, recruitment, data quality, and transparent reporting of the study procedure). 
In the following, we explain the research questions, the study design, questionnaire development, participant recruitment, data quality procedures, and data analysis methods. We show our overall research workflow in Figure~\ref{fig:study_workflow}.

%

\subsection{Research Questions}

The study addresses the following research questions:

\begin{itemize}

\item \textbf{RQ1 (Motivation):} \emph{How do motivations for engaging in vibe coding differ between non-, novice, and professional developers?}

This question examines the reasons individuals adopt vibe coding. Potential motivations include increasing development speed, reducing cognitive effort, supporting learning, or enabling creative exploration. Comparing these motivations across experience levels helps determine whether vibe coding serves different purposes for individuals with different programming backgrounds.

\item \textbf{RQ2 (Experience):} \emph{How do experiences of vibe coding differ between non-, novice, and professional developers?}

This research question focuses on participants’ subjective experiences when engaging in vibe coding. These experiences include perceptions of success, frustration, flow, confidence, and reliance on iterative prompting. Examining these experiences across groups provides insight into how prior programming expertise shapes the practical use of AI-assisted coding.

\item \textbf{RQ3 (Code Quality Perception):} \emph{How do perceptions of AI-generated code quality produced through vibe coding differ between non-, novice, and professional developers?}

This question investigates how participants evaluate the quality of AI-generated code. In particular, it considers perceptions of correctness, maintainability, fragility, and suitability for real-world or production use. Differences across experience levels help reveal how technical expertise influences trust in and evaluation of \aigc{}.

\item \textbf{RQ4 (QA Practices):} \emph{How do QA practices during vibe coding differ between non-, novice, and professional developers?}

This research question examines how participants verify or validate the code generated by AI tools. It includes practices such as testing, code inspection, debugging, reprompting, or delegating verification tasks to the AI system. Comparing these behaviours across experience levels provides insight into how traditional software engineering QA practices are adapted, modified, or bypassed during vibe coding.

\end{itemize}

\subsection{Questionnaire Design}
\label{Meth-Surv-design}
This study employs a quantitative cross-sectional approach, in which data are collected from participants at a single point in time to enable systematic comparison between independent groups \cite{creswell2018research}. Such designs are well-suited for examining group-level differences in motivations, experiences, perceptions of code quality, and QA practices associated with vibe coding (e.g., \cite{bacchelli2013expectations,hilton2016usage}).




The questionnaire was informed by the findings of our recent GLR \cite{fawzy2025vibe}, which identified recurring behavioural themes describing how coders engage in vibe coding practices. These themes provided the conceptual foundation for defining the construct blocks examined in the questionnaire (see Table~\ref{tab:glr_survey_mapping}). Participants were grouped into three independent categories based on self-reported programming experience:
\begin{itemize}
    \item \textbf{Non-developers:} Participants who reported that they do not have formal programming training or professional experience, but have used AI coding tools (e.g., ChatGPT, Copilot) to generate or modify code.
    
    \item \textbf{Novice developers:} Participants who reported having completed tutorials, coursework, or small personal projects involving programming or AI coding tools, but having limited real-world development experience.
    
    \item \textbf{Professional developers:} Participants who reported having professional or academic experience writing, maintaining, or reviewing code as part of their job, research, or formal development work.
\end{itemize}





%
%


\subsection{Questionnaire Development}
\label{method-Quest-Dev}
We used the themes identified in our recent GLR~\cite{fawzy2025vibe} to guide the development of the questionnaire. These themes were organised into four sections corresponding to the research questions: Motivation (RQ1), Experiences (RQ2), Code Quality Perception (RQ3), and QA Practices (RQ4). The questionnaire consisted of the following sections:

\begin{itemize}
\item \textbf{Consent and Definition:}
This section introduced the concept of vibe coding and obtained the participants' consent.

\item \textbf{Experience Classification and Vibe-Coding Style:}
This section collected self-reported programming experience to classify participants into non-developer, novice developer, or professional developer groups. Additional items captured the context of vibe coding use (e.g., personal projects, work-related tasks, and learning activities), the \AItools{} used (e.g., ChatGPT, GitHub Copilot, and Claude Code), the duration and frequency of vibe coding, and the types of projects undertaken (e.g., scripting, web/mobile applications, prototypes, etc.). A \emph{Vibe-Coding Style} subsection was also included to capture participants' preferred workflow characteristics when interacting with \AItools{}.

\item \textbf{Vibe Coding Motivations (RQ1):}
This section of the questionnaire assesses participants' motivation to engage in vibe coding. This includes reasons such as improving development speed, supporting learning, enhancing creativity, or reducing cognitive load. Participants rated statements such as ``I use vibe coding to save time.'' on a five-point scale ranging from \emph{strongly disagree} to \emph{strongly agree}.

\item \textbf{Participants' Experiences (RQ2):}
This section captured participants' experiences while vibe coding, including feelings of success, frustration, flow, reprompting behaviour, and task abandonment. For example, participants rated statements such as ``I often experience a quick flow where everything just works when vibe coding'' and ``I usually reprompt several times to get usable results.''

\item \textbf{Perceived Code Quality (RQ3):}
This section examined participants' perceptions of the quality of \aigc{}, including perceived fragility, maintainability, and suitability for production use. For example, participants are asked to rate statements such as ``Vibe-coded code is often fragile or error-prone'' and ``Vibe-coded code is good enough for demos but not for production use.''

\item \textbf{QA Practices (RQ4):}
This section measured QA practices during vibe coding, including running code without checks, reviewing \aigc{}, re-prompting instead of manual debugging, and delegating QA tasks back to the AI. Both Likert-scale and frequency-based items were used to capture these behaviours. For example, participants rated statements such as ``I run \aigc{} without checking it first'' and reported how often they checked \aigc{} before use using response options ranging from \emph{never} to \emph{always}.

\item \textbf{Demographics:}
The final section collected demographic information from participants, including age range, gender, geographic location, and the highest level of education in a computing-related discipline. These variables were used for descriptive reporting and to characterise the participant population.

\item \textbf{Standalone Open-Ended Questions:}
We included several open-ended questions to gather broader reflections on \aigc{} features, suggestions for improvement, and general feedback on vibe coding practices. These items were presented to all participants and provided a qualitative context that complemented the structured Likert-scale responses.
\end{itemize}

Within each of these sections, every theme was operationalised as a Likert-scale item by simplifying its description (from the original study) into a measurable statement while preserving its original meaning, ensuring alignment between the qualitative synthesis and the quantitative measurement. For example, the theme \emph{``Fast but Flawed''}, described in the GLR as ``useful for quick tasks but unsuitable for production deployment,'' was operationalised as the item \emph{``Code from vibe coding is useful for quick prototypes but often flawed''}. The full mapping between themes and questionnaire items is presented in Table~\ref{tab:glr_survey_mapping}, with detailed theme definitions reported in~\cite{fawzy2025vibe}. The questionnaire was implemented using Qualtrics\footnote{\url{https://www.qualtrics.com/}}. Items were phrased as declarative statements and measured using five-point Likert-type scales (between 1 (Strongly Disagree) and 5 (Strongly Agree)), enabling ordinal comparison across participants and supporting the planned non-parametric group comparison analyses~\cite{field2024discovering,conover1999practical} (see Section~\ref{Meth-QuantStatAnaly}). In addition to Likert-scale items, optional write-in fields were provided after each section to capture responses not covered by the predefined items and were analysed qualitatively as part of construct-coverage validation (see Section~\ref{Meth-ConstructCoverageVer}).

The \emph{Vibe-Coding Style} items were informed by the development paradigm framework proposed in a recent survey of vibe coding models~\cite{ge2025survey}, which highlights the roles of human evaluation, structured development processes, and context management in human--AI coding interaction. The five style themes shown in Table~\ref{tab:q10_Vstyle} map onto these aspects, capturing variation in how coders delegate control to the AI, structure their interaction, and provide contextual guidance.

\renewcommand{\arraystretch}{1.1}

\begin{longtable}{p{5.5cm}
>{\RaggedRight\arraybackslash}p{\dimexpr\linewidth-8cm\relax}
p{1cm}}

\caption{Traceability mapping from themes identified in GLR~\cite{fawzy2025vibe} to items in questionnaire}
\label{tab:glr_survey_mapping} \\

\toprule
\textbf{Theme from GLR} & \textbf{Questionnaire Item} & \textbf{RQ} \\
\midrule
\endfirsthead

\toprule
\textbf{Theme from GLR} & \textbf{Questionnaire Item} & \textbf{RQ} \\
\midrule
\endhead

\bottomrule
\endfoot

\bottomrule
\endlastfoot

\multicolumn{3}{l}{\textbf{RQ1 -- Motivations}} \\

Speed \& Efficiency & I vibe code mainly to save time. & RQ1 \\
Accessibility \& Empowerment & Vibe coding lets me build software I could not have built otherwise. & RQ1 \\
Learning \& Experimentation & I use vibe coding to learn or explore new languages or frameworks. & RQ1 \\
Creative Exploration & I find vibe coding creative and enjoyable, and I often use it for exploration or play. & RQ1 \\
Fast Prototyping & Vibe coding helps me quickly create prototypes or proofs of concept. & RQ1 \\
Reducing Mental Effort & Vibe coding reduces the mental effort required for repetitive or boilerplate coding tasks. & RQ1 \\
Frustration Avoidance & I use vibe coding to avoid tedious or frustrating debugging work. & RQ1 \\
Escape from Complexity & Vibe coding lets me focus on ideas without worrying about complex design or architecture. & RQ1 \\
Curiosity or Play & Sometimes I vibe-code just for fun, curiosity, or experimentation. & RQ1 \\

\midrule
\multicolumn{3}{l}{\textbf{RQ2 -- Experiences}} \\

Instant Success \& Flow & I often experience a quick flow where everything just works when vibe coding. & RQ2 \\
Prompt Struggle \& Iteration & I usually reprompt several times to get usable results. & RQ2 \\
Code Breakdown or Abandonment & I sometimes abandon a task when AI output becomes too messy or buggy. & RQ2 \\
Fun \& Creative Satisfaction & Working this way feels fun and creatively satisfying. & RQ2 \\
AI Hallucinations & \aigc{} sometimes looks correct but behaves incorrectly when run. & RQ2 \\
Confusion or Misunderstanding & Sometimes I feel confused about what the AI understood from my prompt. & RQ2 \\

\midrule
\multicolumn{3}{l}{\textbf{RQ3 -- Code Quality Perception}} \\

Fast but Flawed & Code from vibe coding is useful for quick prototypes, but often flawed. & RQ3 \\
Fragile or Error-Prone & Vibe-coded code is often fragile or error-prone. & RQ3 \\
Sloppy or Low Maintainability & Vibe-coded code is hard to maintain or extend later. & RQ3 \\
Prototype-Ready Only & Vibe-coded code is good enough for demos but not for production use. & RQ3 \\
High Quality \& Clean & \aigc{} is sometimes well-structured and clean. & RQ3 \\
Misleading Confidence & \aigc{} looks correct but hides deeper problems. & RQ3 \\

\midrule
\multicolumn{3}{l}{\textbf{RQ4 -- QA Practices}} \\

Skipped QA & I run \aigc{} without checking it first. & RQ4 \\
Manual Testing or Edits & I test or review \aigc{} carefully before using it. & RQ4 \\
Uncritical Trust & I trust \aigc{} even when I don't fully understand it. & RQ4 \\
Delegated QA to AI & I ask the AI to find or fix its own mistakes. & RQ4 \\
Reprompting Instead of Debugging & If an error occurs, I reprompt instead of manually debugging. & RQ4 \\
Run-and-See Validation & I usually run the code once to see if it works. & RQ4 \\
QA Breakdown or Confusion & Sometimes I give up on debugging because I can't understand the AI's code. & RQ4 \\
\end{longtable}

\begin{table}[H]
\centering
\caption{Vibe-Coding Style}
\label{tab:q10_Vstyle}
\begin{tabularx}{\linewidth}{p{4cm} X}
\toprule
\textbf{Style Theme} & \textbf{Questionnaire Item} \\
\midrule
AI-led Generation & I let the AI generate most or all of the code with little or no intervention. \\
Interactive Dialogue & I interact iteratively with the AI, prompting, reviewing, and refining code through dialogue. \\
Structured Planning & I outline or plan tasks before prompting the AI, using a structured or step-based approach. \\
Verification-Oriented Prompting & I define tests, checks, or verification steps before or alongside prompting the AI. \\
Rich Context Provision & I provide detailed context such as data, examples, or external files to guide the AI's coding. \\
\bottomrule
\end{tabularx}
\end{table}

\noindent \textbf{Required and Optional Items.}
Core construct blocks (experience level classification and the main Likert-scale items corresponding to RQ1--RQ4) were configured as mandatory in Qualtrics to ensure complete data for between-group statistical comparisons. Standalone open-ended question, write-in fields following each Likert block, and demographic questions were optional to reduce participant burden and avoid requiring responses in non-essential fields.

\noindent \textbf{Ethical approval.} The survey was approved by the university ethics committee. Participation was voluntary and anonymous, and participants could ignore/leave out any optional question. The first author developed the initial draft of the questionnaire and refined it through repeated discussions with the other co-authors, resolving disagreements by consensus.

\subsection{Construct Coverage Verification}
\label{Meth-ConstructCoverageVer}

To assess whether the questionnaire provided adequate thematic coverage, we examined the optional write-in responses associated with each Likert-scale construct block. 
These questions allowed participants to describe their vibe coding style, motivations, experiences, perceptions of \aigc{}, or quality-assurance (QA) practices in their own words when the predefined Likert statements did not fully reflect their perspectives. A total of \writein{} write-in responses were collected. Each response was systematically compared with the conceptual dimensions operationalised in the corresponding Likert-scale constructs. 
The responses were manually reviewed by the first author and checked by a co-author. 
Discrepancies were discussed and resolved through consensus, indicating that none of the responses introduced a new conceptual dimension beyond those represented in the questionnaire Likert items. Instead, many responses mapped directly to existing constructs. 
For example, one participant described their workflow as: ``I usually work iteratively with the AI by describing what I want, reviewing the generated code, running it, and making small adjustments as needed'', which corresponds to the Likert item: ``I interact iteratively with the AI, prompting, reviewing, and refining code through dialogue''. Other responses provided contextual descriptions consistent with the predefined themes. For instance, a participant noted that ``\aigc{} can look clean and correct at first glance, but deeper issues sometimes appear when running or extending it'', which aligns with the Likert item: ``Vibe-coded code is hard to maintain or extend later''.
These observations provide additional evidence of construct coverage and support the survey instrument's content validity, suggesting that the questionnaire adequately captured the conceptual scope of vibe coding practices examined in this study.

\subsection{Target Population}
The target population consisted of vibe coders aged 18 years or older who had used \AItools{} for vibe coding. This included non-developers with no formal programming background, novice developers with limited experience, and professional developers with professional or academic coding experience. Including all three groups enabled a systematic examination of how vibe coding practices vary with experience.

\subsection{Determining Sample Size}
We employed a power analysis to determine the minimum required sample size to detect differences in vibe coding practices across the three independent experience groups. We used a one-way between-groups ANOVA to estimate the minimum required sample size for group comparisons \cite{cohen2013statistical}. Although Likert-type responses are ordinal, they are commonly treated as approximately continuous in power analyses, thereby enabling ANOVA-based sample size estimation \cite{field2024discovering}. Furthermore, ANOVA-based power calculations tend to slightly overestimate the required sample size when the final analysis uses non-parametric tests, making this a conservative planning approach \cite{conover1999practical}. The power analysis assumed a medium effect size (Cohen’s $f = 0.25$) across three groups (non-developers, novice developers, and professional developers), with a significance level of $\alpha = 0.05$ and desired power of $0.80$, representing the smallest group difference the study was designed to detect.




The analysis was conducted in R using the ``pwr'' package (pwr.anova.test). Under these assumptions, the minimum required sample size was \RequiredSampleSize{} participants (53 per group). A total of \RecivedResponses{} responses were collected. After applying the predefined data-quality protocol (see Section \ref{Meth-DataQ-protcol}), the analysed sample comprised \AnalysedSampleSize{} participants (54 per group), exceeding the minimum threshold.


To the best of our knowledge, no prior quantitative survey studies estimate effect sizes for differences in vibe coding practices across experience levels. Following standard guidance, a medium effect size (Cohen’s $f = 0.25$) was assumed for planning purposes \cite{cohen2013statistical}. This assumption defines the smallest effect the study was designed to reliably detect and does not imply that the observed effects were expected to be of medium magnitude.


\subsection{Pilot Study}
\label{Meth-Pilot}
To validate the clarity and suitability of the questionnaire, we conducted a pilot run before its full deployment. From the pool of participants (see Section \ref{Meth-Data-collection}) who reported engaging in vibe coding, we randomly selected 9 candidates (3 from each experience group) and invited them to participate in the pilot study. The pilot run collected feedback on whether the questionnaire questions were clear, well-worded, and appropriately captured the intended constructs (motivations, experiences, perceptions of \aigc{}, and QA practices). Participants were also invited to comment on potential issues, such as ambiguity, bias, or gaps. \Space{The pilot study provided useful feedback. For example, one participant commented, ``The questions were clear and there were no particular issues, but the answer choices for the question about the AI tools used seemed a bit biased''. Another participant noted that all questions were clear, except for a minor issue with the structure of the answers to one question. The comments we received indicated minor issues with the response scale balance and potential bias in the answer options.}

Based on the pilot survey feedback, we refined the questionnaire by adjusting response scales, clarifying wording, and removing sources of bias. Using the revised version, we deployed the final questionnaire.

\subsection{Data Collection}
\label{Meth-Data-collection}
We used Prolific \footnote{\url{https://www.prolific.com/}} to recruit human participants, with recruitment conducted in multiple stages. First, we created and published two screening questions on Prolific to identify potential participants who (1) had experience with vibe coding and (2) belonged to one of the three target groups: non-software developers, novice developers, or professional developers. This screening process resulted in a pool of \pool{} candidate participants, consisting of \poolnoneDev{} non-software developers, \poolNovice{} novice developers, and \poolProf{} professional developers. Of these, \poolEng{} reported engaging in vibe coding, while \poolNotEng{} reported no prior experience with vibe coding. From the pool of eligible vibe coding participants, we invited \poolEngInvitated{} participants. We received responses from \RecivedResponses{} participants. 
The experience-classification question was administered twice, at screening (5 January 2026) and again at the start of the main questionnaire (3 February 2026), approximately four weeks apart. The main questionnaire response was used for group assignment. Of the \AnalysedSampleSize{} participants in the sample, \matchedScreening{} could be matched to a screening response via their Prolific identifier. Among these, \retestSame{} (\retestSamePct{}\%) gave the same classification on both occasions, while \retestChanged{} (\retestChangedPct{}\%) reported a different one, with \retestToMiddle{} of the changes drifting toward the middle ``novice'' category and only one participant moving between non-developer and professional developer. Given the four-week gap, we did not exclude participants whose responses differed.

\subsection{Data Quality and Reliability Checks}
\label{Meth-DataQ-protcol}
Before conducting our analysis, the questionnaire responses were screened using a structured data quality protocol 
consistent with established recommendations for detecting inattentive responding (i.e., answering without carefully reading questionnaire items) and insufficient-effort responding (i.e., responses provided with minimal cognitive engagement) in online questionnaires \cite{Leiner2019,MeadeCraig2012,Curran2016}. All screening procedures were pre-specified and applied uniformly across experience groups to minimise the risk of systematic bias.
The screening protocol was implemented in predefined stages. Steps~1--3 constituted the sequential exclusion criteria: responses failing a given criterion were removed before proceeding to the next stage. Steps~4 and~5 served as additional validation checks and did not independently result in participant exclusion.\\

\noindent\emph{\textbf{Step 1}: Completion filtering}.
Only fully completed responses were retained for analysis. Responses that were started but not finished were excluded. In total, 13 incomplete responses were removed.\\

\noindent\emph{\textbf{Step 2}: Time-based cognitive plausibility}.
We also examined completion times to detect implausibly rapid responses. The median completion time was 9.8 mins ($M = 11.9$, $SD = 7.3$ mins). Following best-practice guidance for detecting inattentive and insufficient-effort responses in online questionnaires \cite{Leiner2019,MeadeCraig2012,Curran2016,huang2015insufficient,desimone2015best}, we derived a conservative lower-bound estimate of the minimal attentive completion time based on the questionnaire's structure. This estimate accounted for the time required to read the consent and introductory definition, respond to the 33 mandatory items, and provide minimal engagement with the demographic and optional open-ended questions. Under these assumptions, the minimum plausible attentive completion time was estimated at approximately 4 mins. Responses completed in under four mins were therefore considered inconsistent with attentive responding. The 4 mins cutoff corresponds to approximately 41\% of the observed median completion time (9.8 mins) and falls below the empirical lower tail of the completion-time distribution (5th percentile $\approx$ 5 mins), indicating that it lies well below typical completion durations. Responses below this threshold were classified as implausibly rapid, resulting in the exclusion of 6 responses.\\


\noindent\emph{\textbf{Step 3}: Response-pattern diagnostics.}
We applied response-pattern diagnostics across all Likert-scale items 
to detect straight-lining and invariant responding. Straight-lining refers to selecting the same response option for all Likert items\Space{ (e.g., selecting 4, 4, 4, 4 across all 33 items)}. Invariant responding refers to zero variance (standard deviation = 0) across all Likert responses.
Such patterns are commonly associated with satisficing behaviour, whereby respondents provide minimally effortful answers rather than evaluating each item carefully \cite{MeadeCraig2012,Curran2016}. No responses met these criteria; therefore, no additional exclusions were applied at this stage.\\

\noindent\emph{\textbf{Step 4}: Open-ended engagement validation.}
All open-ended responses were screened for substantive engagement (interpretable content directly related to the question prompt). For example, the question ``What features would make \aigc{} (vibe coding output) easier and safer for you to use or trust?'' received meaningful responses from all 162 participants.
Empty or clearly non-substantive entries (e.g., random characters or irrelevant text) were identified and flagged in the dataset. This step served as an additional engagement validation layer and did not result in participant removal.\\

\noindent\emph{\textbf{Step 5}: Robustness verification.}
As a robustness check, we repeated the full analysis using a stricter completion-time threshold of five minutes, yielding a reduced sample of N = 157. All inferential tests (Kruskal--Wallis with Benjamini--Hochberg correction, effect sizes, and Dunn's post-hoc comparisons) were re-run on this sample. The sensitivity analysis preserved the same pattern of statistically significant differences, effect-size magnitudes, and central tendencies across all four research questions: experience-level differences remained concentrated in interaction style, motivations (RQ1), and QA practices (RQ4), with no significant differences in experiences (RQ2) or perceived code quality (RQ3). No previously significant result disappeared, and no new significant result emerged, indicating that the study's conclusions are not sensitive to the response-time threshold.

Overall, 19 responses (10.5\% of the original sample) were excluded during data screening, yielding a final analytical sample of 162 participants. These procedures were implemented to reduce measurement error arising from inattentive or insufficient-effort responding while preserving valid responses and minimising the risk of systematic exclusion bias.












\subsection{Data Analysis}
\label{Meth-DataAnalysis}
We analysed the survey responses using descriptive and inferential statistics. Descriptive statistics summarised response distributions across experience groups, while inferential tests examined whether statistically significant differences existed between groups.
Most questionnaire items were measured on 5-point Likert scales and were therefore analysed using non-parametric statistical methods. Questions capturing categorical responses (e.g., application context and project types) were analysed using chi-square tests of independence.

\subsubsection{Quantitative Statistical Analysis}
\label{Meth-QuantStatAnaly}

\noindent \textbf{Descriptive Analysis.}
Before conducting inferential statistical tests, we summarised survey responses using descriptive statistics for each experience group (non-developers, novices, and professionals). Because the questionnaire items were measured on a 5-point Likert scale, responses were treated as ordinal data. For each Likert item, the median and interquartile range (IQR) were calculated for each group. 
The IQR (the range between the 25th and 75th percentiles) reflects the dispersion of responses within the group. These statistics are appropriate for ordinal data because they do not assume equal distances between scale points \cite{field2024discovering}. Descriptive statistics were used to characterise overall patterns of agreement and illustrate how responses differed across experience groups before formal statistical testing. In particular, comparing medians across groups helps illustrate potential differences in central tendency prior to formal statistical testing.


\noindent \textbf{Inferential Statistical Tests.}
For each questionnaire item, differences between the three independent experience groups (non-developers, novices, and professionals) were assessed using the Kruskal--Wallis $H$ test \cite{field2024discovering}. The Kruskal--Wallis $H$ test was selected because the questionnaire responses were measured on ordinal Likert-type scales and the study compares more than two independent groups (non-developers, novices, and professionals). As a rank-based non-parametric test, it does not assume normally distributed data and is therefore appropriate for analysing ordinal and potentially skewed responses \cite{conover1999practical}. 
In addition to statistical significance testing, effect sizes were calculated using epsilon-squared ($\varepsilon^2$). This measure estimates the proportion of variability in ranked responses that can be attributed to differences between groups. 
Epsilon-squared was selected because it is well-suited to rank-based non-parametric tests and provides an interpretable estimate of effect magnitude for Kruskal-Wallis comparisons.

\noindent \textbf{Multiple comparisons correction.}
We also carried out a significant correction for multiple testing. For example, RQ1 (motivations) consisted of several individual Likert-scale items, each analysed separately using the Kruskal-Wallis test. Conducting multiple statistical tests within the same research question increases the probability of identifying significant results purely by chance. To address this, $p$-values were adjusted using the Benjamini--Hochberg false discovery rate (FDR) procedure \cite{benjamini1995controlling}. The Benjamini--Hochberg method controls the expected proportion of false discoveries among rejected hypotheses while maintaining higher statistical power than family-wise error rate corrections. FDR-based approaches are widely recommended in behavioural and empirical research because they are more likely to detect real differences (i.e., have higher statistical power) than more conservative corrections such as Bonferroni, which can miss real effects (Type II errors) 
\cite{groppe2011mass,verhoeven2005implementing}. For questionnaire items that remained statistically significant after the Benjamini--Hochberg adjustment, Dunn’s post-hoc pairwise comparisons were conducted to determine which specific experience groups differed. 
To control the increased risk of false-positive findings when performing multiple pairwise comparisons, Holm’s correction was applied to the Dunn test $p$-values \cite{holm1979procedure}. Holm’s sequential procedure controls the family-wise error rate while being less conservative and more powerful than the traditional Bonferroni correction.

\noindent \textbf{Categorical Comparisons.}
We analysed categorical questionnaire questions using inferential statistical tests appropriate for nominal data. Two questionnaire questions captured categorical variables rather than Likert-scale responses: (\textit{Vibe-Coding Application Context} and \textit{Types of Projects or Tasks Performed with Vibe Coding}). 
 For these items, differences between experience groups were analysed using Pearson’s chi-square ($\chi^2$) test of independence. 
When significant associations were observed, effect sizes were reported using Cramér’s $V$.



\noindent\textbf{Regression Analysis.}
We conducted a follow-up analysis to examine which dimensions of experience drive differences in QA (RQ4). We modelled the three significant QA outcomes using ordinal logistic regression~\cite{mccullagh1980regression,agresti2010analysis}, with experience group as the main predictor (non-developers as the reference category) and three usage-related controls: weekly vibe coding hours, adoption duration, and weekly independent (non-vibe) coding activity. These controls were entered as ordered numeric scores. The proportional-odds assumption was verified for all models using the Brant test~\cite{brant1990assessing}. Multicollinearity was assessed using generalised variance inflation factors (GVIF)~\cite{fox1992generalized}.

\subsubsection{Qualitative Analysis}
We conducted a qualitative analysis on responses to the open-ended survey questions 
, which asked participants: \textit{``What features would make \aigc{} (vibe coding output) easier and safer for you to use or trust?''} and providing any additional comments. 

Responses were analysed using a lightweight thematic analysis \cite{braun2006thematic} to identify recurring patterns in participants’ descriptions. First, all responses were read in full to familiarise the researchers with the data. The first author then generated initial codes representing meaningful units of text (e.g., requests for better documentation, improved testing support, or stronger security guarantees). Related codes were subsequently grouped into broader themes that captured common expectations for improving the usability and trustworthiness of \aigc{}.
Themes were iteratively reviewed and refined by the first author and discussed with other co-authors to ensure that coded responses were consistently interpreted and that the resulting themes accurately reflected the dataset. The qualitative themes were then used to complement the quantitative findings by providing concrete examples of how participants described desired improvements to vibe coding outputs.

\section{Results}
\label{surveyresults}

\subsection{Sample Overview}
After applying the predefined data quality protocol (Section~\ref{Meth-DataQ-protcol}), the final analytical sample comprised \AnalysedSampleSize{} participants, evenly distributed across three independent experience groups: 54 non-software developers, novice developers, and experienced developers, each.

\subsubsection{Participant Demographics}
\label{Results-Demographics}
\textcolor{black}{The three experience groups were broadly comparable in age but differed more in gender composition and educational background. Participants were predominantly aged 25--44 ($n=116$, 71.6\%), with 25--34 the largest category ($n=80$, 49.4\%). Most participants self-identified as men ($n=101$, 62.3\%) and women ($n=59$, 36.4\%). The sample reflects broad international participation across more than 20 countries, the largest groups being South Africa ($n=29$, 17.9\%), the United Kingdom and the United States ($n=16$, 9.9\% each), Portugal ($n=13$, 8.0\%), and Brazil ($n=12$, 7.4\%). Educational background varied across the sample: the most common response was a bachelor's or honours degree ($n=71$, 43.8\%), followed by a master's degree or higher ($n=24$, 14.8\%).}

\subsubsection{Vibe Coding Engagement}
\begin{figure}[t]
\centering
\includegraphics[width=\linewidth]{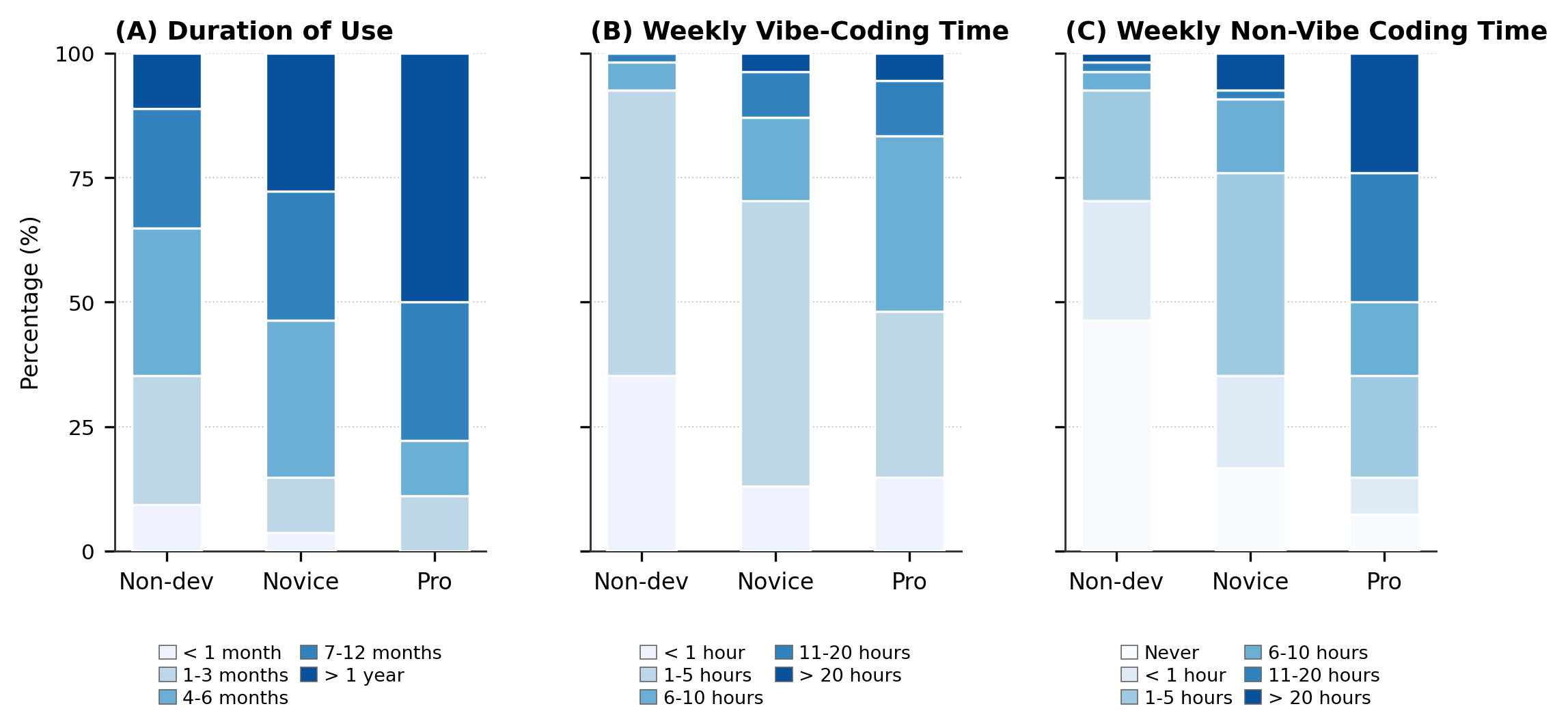}
\caption{Vibe coding engagement across experience groups.}
\label{fig:vibe-engagement}
\end{figure}

\textcolor{black}{Participants reported sustained but heterogeneous engagement with vibe coding, and all three engagement measures differed significantly across experience groups (Figure~\ref{fig:vibe-engagement}). Adoption duration varied by experience ($\chi^2(8)=30.06$, $p<.001$, $V=.305$): experienced developers were the most likely to report long-term use (50\% reporting more than one year), whereas non-developers were more commonly recent adopters. Weekly vibe coding intensity showed a similar pattern ($\chi^2(8)=30.99$, $p<.001$, $V=.309$), with non-developers concentrated in lower-usage categories and experienced developers reporting heavier weekly use. The biggest difference was in coding activity \emph{without} vibe coding ($\chi^2(10)=66.73$, $p<.001$, $V=.454$): non-developers most often reported never coding without vibe coding (46.3\%), indicating high reliance on AI-assisted workflows, whereas roughly half of experienced developers reported 11 or more hours of independent coding per week. Full distributions are shown in Figure~\ref{fig:vibe-engagement}.}

\subsubsection{Vibe Coding Application Context}
\label{Res-VCappContex}
\textcolor{black}{Across the sample, vibe coding was most commonly used for personal projects (62.3\%) and hobby or creative exploration (51.2\%), followed by work-related use (42.6\%) and education (30.9\%) (Table~\ref{tab:q4_context}). Two contexts differed significantly by experience: work-related use rose sharply with experience ($\chi^2(2)=22.42$, $p<.001$, $V=.37$), reported by 68.5\% of experienced developers but only 27.8\% of non-developers, whereas hobby use showed the opposite trend ($\chi^2(2)=9.69$, $p=.008$, $V=.25$), highest among non-developers (64.8\%) and lowest among experienced developers (35.2\%)}. 

\textcolor{black}{By task type, vibe coding was most frequently used for small scripts or automation (62.3\%), creative projects (53.7\%), and data analysis (40.1\%) (Table~\ref{tab:q9_tasks}). Three task types differed significantly by experience. Assignments and learning tasks were most common among novices ($\chi^2(2)=7.09$, $p=.029$, $V=.21$), while work-related production tasks and prototypes or demos were more common among experienced developers ($\chi^2(2)=6.63$, $p=.036$, $V=.20$ and $\chi^2(2)=6.39$, $p=.041$, $V=.20$, respectively).
In terms of tools, ChatGPT was the most widely used across all groups (75.3\%), followed by Gemini (39.5\%) and GitHub Copilot (24.1\%) (Table~\ref{tab:q5_assistants}). Usage rates were broadly similar across experience groups; the only significant difference was Claude ($\chi^2(2)=9.99$, $p=.007$, $V=.25$), which was used more by experienced developers (27.8\%) than by novices (14.8\%) or non-developers (5.6\%).}

\begin{table}[t]
\centering
\caption{Vibe coding context by experience groups (\%)}
\label{tab:q4_context}
\begin{tabular}{lccc}
\toprule
\textbf{Context} & \textbf{Non-dev} & \textbf{Novice} & \textbf{Pro} \\
\midrule
\textbf{Work (job / freelance)} & \textbf{27.8} & \textbf{31.5} & \textbf{68.5} \\
Personal Projects & 55.6 & 72.2 & 59.3 \\
\textbf{Hobby / creative} & \textbf{64.8} & \textbf{53.7} & \textbf{35.2} \\
Education & 20.4 & 38.9 & 33.3 \\
\bottomrule
\end{tabular}
\end{table}

\begin{table}[t]
\centering
\caption{Vibe coding task types by experience groups (\%)}
\label{tab:q9_tasks}
\begin{tabular}{lccc}
\toprule
\textbf{Task Type} & \textbf{Non-dev} & \textbf{Novice} & \textbf{Pro} \\
\midrule
Small scripts / automation & 51.9 & 64.8 & 70.4 \\
Creative / hobby projects & 59.3 & 50.0 & 51.9 \\
Data analysis / visualisation & 31.5 & 44.4 & 44.4 \\
\textbf{Assignments / learning} & \textbf{33.3} & \textbf{50.0} & \textbf{25.9} \\
Web / mobile apps & 31.5 & 31.5 & 37.0 \\
\textbf{Work-related production} & \textbf{16.7} & \textbf{14.8} & \textbf{33.3} \\
\textbf{Prototypes / demos} & \textbf{9.3} & \textbf{24.1} & \textbf{27.8} \\
\bottomrule
\end{tabular}
\end{table}

\begin{table}[t]
\centering
\caption{AI coding assistants used by experience groups (\%)}
\label{tab:q5_assistants}
\begin{tabular}{lccc}
\toprule
\textbf{AI Assistant} & \textbf{Non-dev} & \textbf{Novice} & \textbf{Pro} \\
\midrule
ChatGPT & 77.8 & 68.5 & 79.6 \\
Gemini & 35.2 & 40.7 & 42.6 \\
GitHub Copilot & 18.5 & 25.9 & 27.8 \\
Claude & 5.6 & 14.8 & 27.8 \\
Cursor & 1.9 & 13.0 & 7.4 \\
DeepSeek & 1.9 & 1.9 & 1.9 \\
Grok & 3.7 & 3.7 & 3.7 \\
\bottomrule
\end{tabular}
\end{table}

\subsection{Vibe Coding Style}

\textcolor{black}{Beyond the four research questions, we also examined how participants operationalise vibe coding through their \emph{interaction style} (see Section~\ref{method-Quest-Dev}), a complementary dimension not tied to a specific research question. It is measured through five style themes (Table~\ref{tab:q10_Vstyle}):}

\begin{figure}[t]
    \centering
    \includegraphics[width=\linewidth]{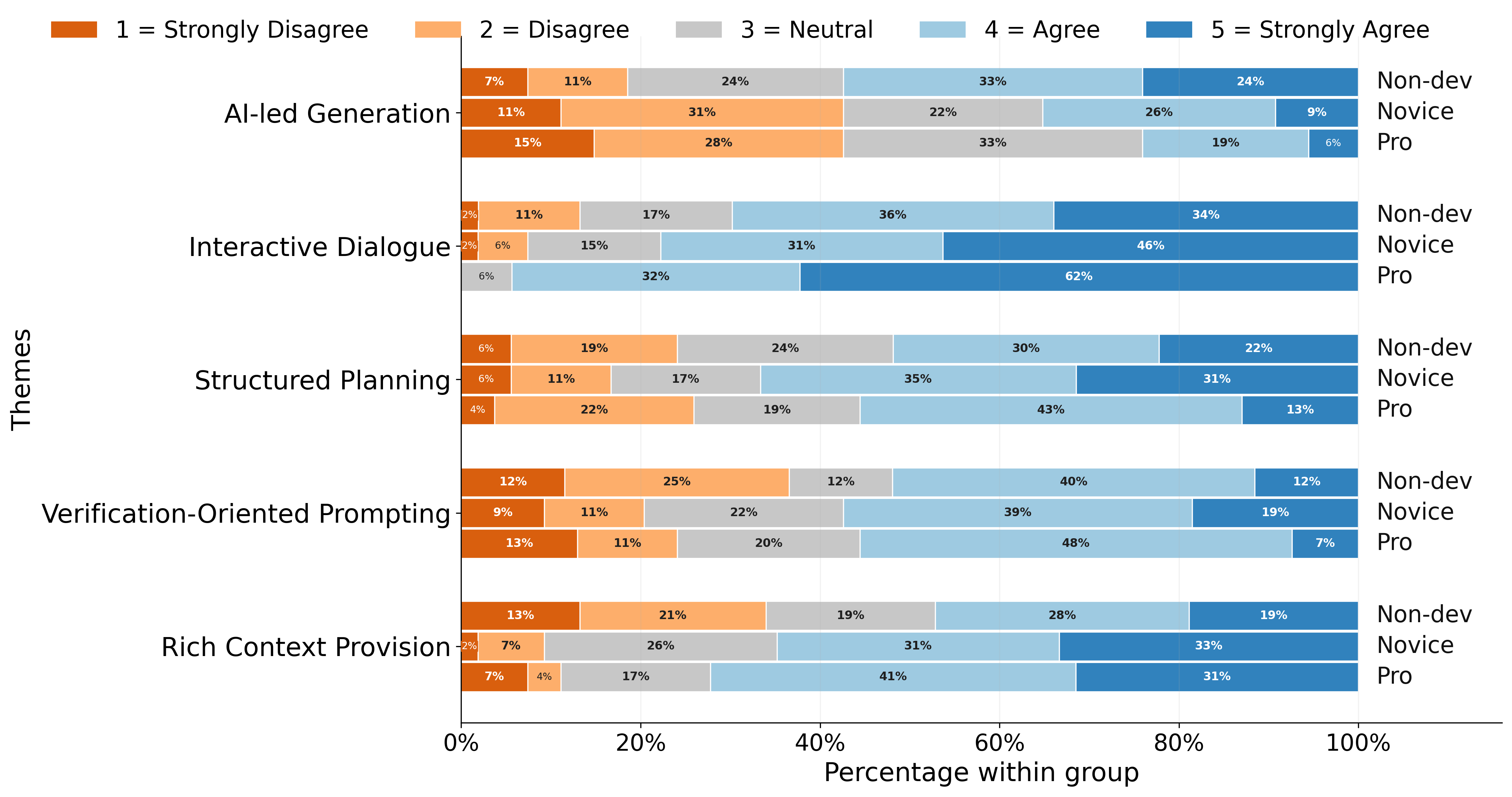}
    \caption{Distribution of vibe coding style responses across groups.}
    \label{fig:q10-style}
\end{figure}

(1)~\textit{AI-led generation} (delegating code generation 
to the \AItools{}), (2)~\textit{interactive dialogue} (iterative prompting and 
refinement), (3)~\textit{structured planning} (pre-planning tasks before prompting), 
(4)~\textit{verification-oriented prompting} (defining tests or checks), and 
(5)~\textit{rich context provision} (supplying detailed input such as data or examples). 
Figure~\ref{fig:q10-style} shows the distribution of responses across experience groups, 
while Table~\ref{tab:q10_style} summarises medians and interquartile ranges. Overall, 
responses were centred around agreement ($\text{Mdn} \approx 4$), indicating that these 
practices are commonly adopted across all groups. Statistically significant differences 
were identified for three themes: \textit{AI-led Generation} 
($p_{\text{BH}}=.004$, $\varepsilon^2=.079$), \textit{Interactive Dialogue} 
($p_{\text{BH}}=.005$, $\varepsilon^2=.067$), and \textit{Rich Context 
Provision} ($p_{\text{BH}}=.015$, $\varepsilon^2=.047$). As shown in Table~\ref{tab:pairwise_all}, non-developers report higher agreement than both novices and professionals on letting the \AItools{} generate most or all of the code with little or no intervention (\textbf{\textit{AI-led Generation}}), while novices and professionals did not differ significantly from each other. For 
\textbf{\textit{Interactive Dialogue}} (capturing iterative prompting, reviewing, and 
refining code through dialogue), professionals report higher agreement than 
non-developers ($p_{\text{adj}}=.001$), with no significant difference between novices 
and either group. For \textbf{\textit{Rich Context Provision}} (reflecting the 
provision of detailed context such as data, examples, or external files to guide the 
\AItools{}), non-developers report lower agreement than both novices and professionals 
(both $p_{\text{adj}}=.019$), with no significant difference between novices and 
professionals.
Structured planning and verification-oriented prompting did not differ significantly 
across groups ($p_{\text{BH}} \geq .177$; Table~\ref{tab:q10_style}), with negligible 
effect sizes ($\varepsilon^2 \leq .012$). Median responses were consistent across groups 
($\text{Mdn} \approx 4$), indicating that these practices are relatively stable 
regardless of experience level.

\noindent\textbf{Overall Pattern.}
Experience level shapes how participants operationalise vibe coding, with differences 
concentrated in three dimensions. Non-developers exhibit greater reliance on AI-led 
generation, reflecting a more passive engagement with the \AItools{}. In contrast, 
professionals engage more actively through iterative dialogue and the provision of richer 
contextual input. Structured planning and verification-oriented prompting remain stable 
across groups, suggesting that these aspects of vibe coding behaviour are broadly shared 
regardless of experience. Together, these results indicate a shift from passive to more 
interactive and context-rich engagement as experience increases, reflecting differences in 
how participants guide, refine, and control \aigc{}.

\begin{table}[t]
\centering
\caption{Vibe Coding Style across experience groups}
\label{tab:q10_style}
\begin{tabular}{lccccc}
\toprule
\textbf{Theme}
& \textbf{Non-dev} & \textbf{Novice} & \textbf{Pro}
& \textbf{$p_{BH}$} & \textbf{$\varepsilon^2$} \\
\midrule

\textbf{AI-led Generation}
& \textbf{4 (1)} & \textbf{3 (2)} & \textbf{3 (1)}
& $\mathbf{.004}$ & \textbf{.079} \\

\textbf{Interactive Dialogue}
& \textbf{4 (2)} & \textbf{4 (1)} & \textbf{5 (1)}
& $\mathbf{.005}$ & \textbf{.067} \\

Structured Planning
& 4 (1) & 4 (2) & 4 (1.75)
& .177 & .012 \\

Verification-Oriented Prompting
& 4 (2) & 4 (1) & 4 (1)
& .444 & .000 \\

\textbf{Rich Context Provision}
& \textbf{3 (2)} & \textbf{4 (2)} & \textbf{4 (2)}
& $\mathbf{.015}$ & \textbf{.047} \\

\bottomrule
\end{tabular}
\end{table}

\begin{table}[t]
\centering
\caption{Pairwise comparisons for all statistically significant themes. 
}
\label{tab:pairwise_all}
\small
\renewcommand{\arraystretch}{1.15}
\setlength{\tabcolsep}{3pt}
\resizebox{\columnwidth}{!}{%
\begin{tabular}{llccc}
\toprule
\textbf{Section} & \textbf{Theme} & \textbf{\shortstack[c]{Non-dev\\vs Novice}} & \textbf{\shortstack[c]{Non-dev\\vs Pro}} & \textbf{\shortstack[c]{Novice\\vs Pro}} \\
\midrule
\multirow{3}{*}{\shortstack[l]{Vibe Coding\\Style}} & AI-led Generation & $\mathbf{.009}$ \small(Non-dev\,$>$\,Novice) & $\mathbf{< .001}$ \small(Non-dev\,$>$\,Pro) & Not significant \\
 & Interactive Dialogue & Not significant & $\mathbf{.001}$ \small(Pro\,$>$\,Non-dev) & Not significant \\
 & Rich Context Provision & $\mathbf{.019}$ \small(Novice\,$>$\,Non-dev) & $\mathbf{.019}$ \small(Pro\,$>$\,Non-dev) & Not significant \\
\midrule
\multirow{2}{*}{\shortstack[l]{Motivations\\(RQ1)}} & Accessibility \& Empowerment & Not significant & $\mathbf{< .001}$ \small(Non-dev\,$>$\,Pro) & $\mathbf{.004}$ \small(Novice\,$>$\,Pro) \\
 & Learning \& Experimentation & $\mathbf{.002}$ \small(Novice\,$>$\,Non-dev) & Not significant & Not significant \\
\midrule
\multirow{2}{*}{\shortstack[l]{QA Practices\\(RQ4)}} & Reprompting Instead of Debugging & Not significant & $\mathbf{.008}$ \small(Non-dev\,$>$\,Pro) & Not significant \\
 & QA Breakdown or Confusion & $\mathbf{.027}$ \small(Non-dev\,$>$\,Novice) & $\mathbf{.002}$ \small(Non-dev\,$>$\,Pro) & Not significant \\
\bottomrule
\end{tabular}
}
\end{table}

\subsection{Motivations for Vibe Coding (RQ1)}
\begin{figure}[t]
    \centering
    \includegraphics[width=\linewidth]{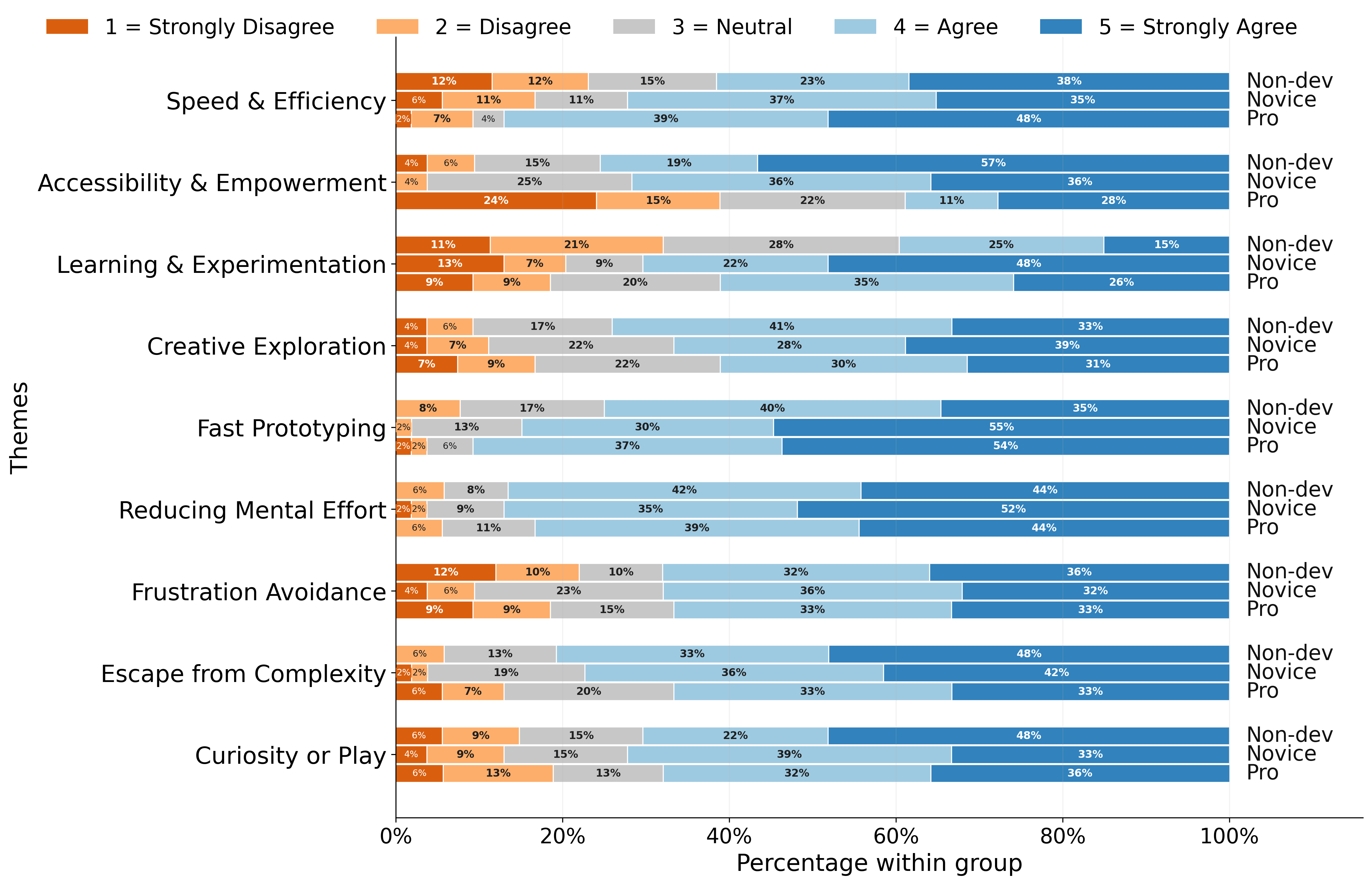}
    \caption{Distribution of motivation responses across experience groups.
    }
    \label{fig:rq1-motivation}
\end{figure}

RQ1 examined participants' motivations for using vibe coding across nine themes. Figure~\ref{fig:rq1-motivation} shows the distribution of responses across experience groups, while Table~\ref{tab:rq1_motivation} summarises medians and interquartile ranges. Overall, motivations were consistently rated around agreement ($\text{Mdn} \approx 4$) across all groups. Statistically significant differences were observed in two themes: \textit{Accessibility \& Empowerment} ($p_{\text{BH}}<.001$, $\varepsilon^2=.110$) and \textit{Learning \& Experimentation} ($p_{\text{BH}}=.015$, $\varepsilon^2=.059$). Table~\ref{tab:pairwise_all} shows that both non-developers and novices report higher agreement than professionals on the item stating that vibe coding lets them build software they could not have built otherwise (\textbf{\textit{Accessibility \& Empowerment}}), with no significant difference between non-developers and novices. Furthermore, non-developers report lower agreement than novices on using vibe coding to learn or explore new languages or frameworks (\textit{\textbf{Learning \& Experimentation}}), with no other significant pairwise differences. All remaining motivation themes did not differ significantly across experience groups 
($p_{\text{BH}} \geq .104$). Median responses were consistently around agreement ($\text{Mdn} \approx 4$), indicating that motivations such as speed and efficiency, creative exploration, reduced mental effort, frustration avoidance, and curiosity are broadly shared across experience levels.

\noindent\textbf{Overall Pattern.}
Motivations for vibe coding are broadly similar across experience levels, with differences limited to specific themes. Accessibility and empowerment are most strongly associated with non-developers, reflecting their role as an enabling mechanism, whereas learning and experimentation are more prominent among novices, indicating that they use vibe coding to develop programming understanding and experiment with code behaviour. In contrast, motivations related to speed, creativity, and reduced effort are consistently reported 
across groups, indicating that these motivations are largely independent of prior programming experience.

\begin{table}[t]
\centering
\caption{Motivation themes for vibe coding across experience groups.}
\label{tab:rq1_motivation}
\begin{tabular}{lccccc}
\toprule
\textbf{Theme}
& \textbf{Non-dev} & \textbf{Novice} & \textbf{Pro}
& \textbf{$p_{BH}$} & \textbf{$\varepsilon^2$} \\
\midrule
Speed \& Efficiency
& 4 (2) & 4 (2) & 4 (1)
& .884 & .000 \\

\textbf{Accessibility \& Empowerment}
& \textbf{5 (1)} & \textbf{4 (2)} & \textbf{3 (3)}
& $\mathbf{<.001}$ & \textbf{.110} \\

\textbf{Learning \& Experimentation}
& \textbf{3 (2)} & \textbf{4 (2)} & \textbf{4 (1.75)}
& $\mathbf{.015}$ & \textbf{.059} \\

Creative Exploration
& 4 (1.75) & 4 (2) & 4 (2)
& .884 & .000 \\

Fast Prototyping
& 4 (1.25) & 5 (1) & 5 (1)
& .104 & .018 \\

Reducing Mental Effort
& 4 (1) & 5 (1) & 4 (1)
& .177 & .012 \\

Frustration Avoidance
& 4 (2) & 4 (2) & 4 (2)
& .884 & .000 \\

Escape from Complexity
& 4 (1) & 4 (1) & 4 (2)
& .884 & .000 \\

Curiosity or Play
& 4 (2) & 4 (2) & 4 (2)
& .884 & .000 \\
\bottomrule
\end{tabular}
\end{table}

\subsection{Experience with Vibe Coding (RQ2)}
\begin{figure}[t]
    \centering
    \includegraphics[width=\linewidth]{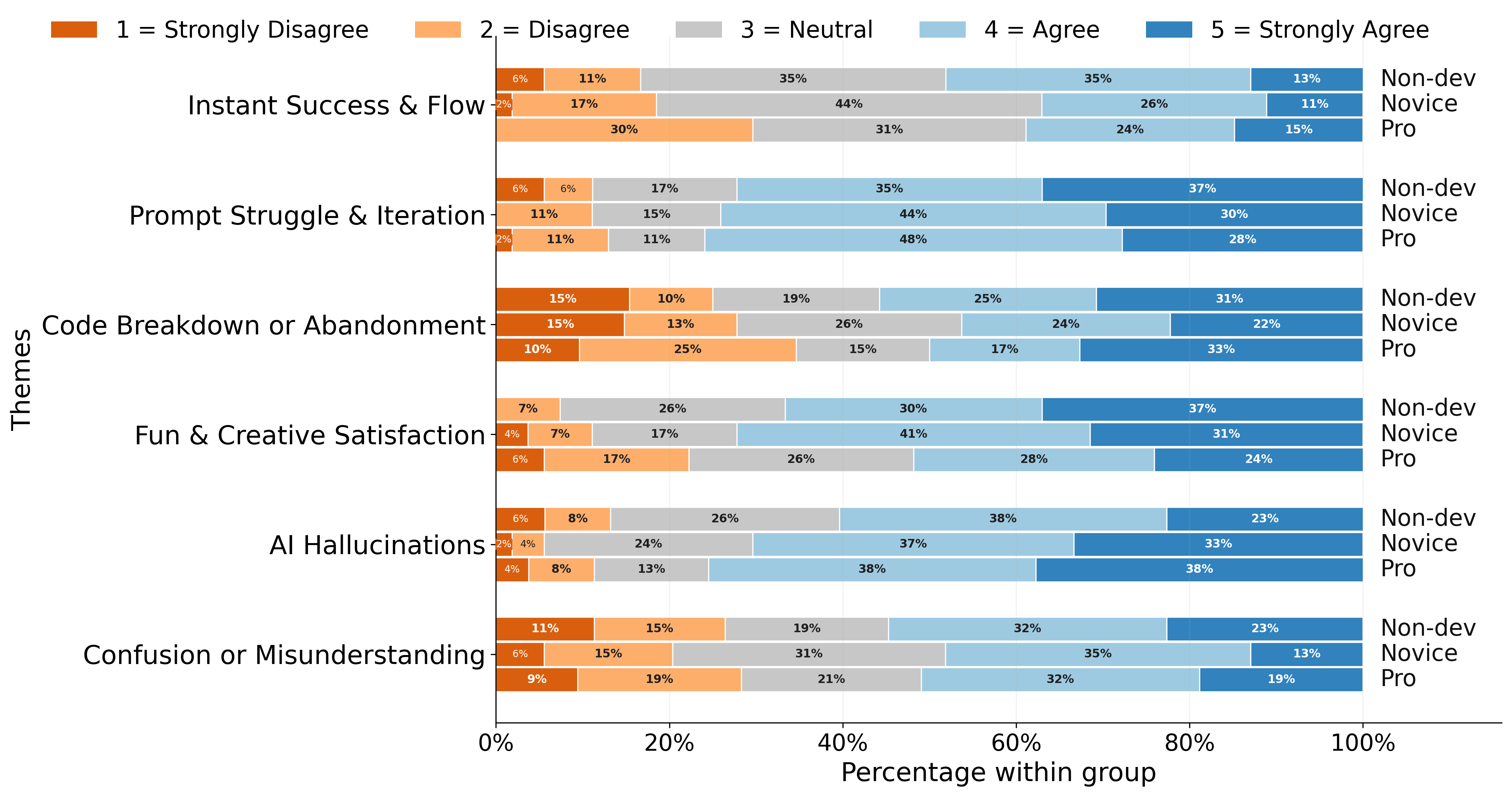}
    \caption{Distribution of experience responses across groups.}
    \label{fig:rq2-experience}
\end{figure}

RQ2 examined participants' experiences during vibe coding, including both positive states 
(e.g., flow and creative satisfaction) and process-related challenges (e.g., iteration, 
hallucinations, and confusion). Figure~\ref{fig:rq2-experience} shows the distribution of 
responses across experience groups, while Table~\ref{tab:rq2_experience} summarises medians and interquartile ranges. No statistically significant differences were observed 
across experience levels for any theme ($p_{\text{BH}} \geq .437$, $\varepsilon^2 \leq 
.020$), and responses were broadly consistent across groups, ranging between neutral and 
agreement ($\text{Mdn} = 3$--$4$). Positive experiences were similarly reported across all groups. Participants moderately agreed that vibe coding is fun and creatively satisfying, and that it occasionally produces a quick flow where everything just works. Challenges were equally shared. 
Participants consistently agreed that reprompting is a common part of the process, that 
\aigc{} can appear correct but behave incorrectly when executed, and that they sometimes 
feel uncertain about what the \AItools{} understood from their prompt. Task abandonment 
when \aigc{} becomes too messy or buggy was also reported at moderate levels across all 
groups.

\noindent\textbf{Overall Pattern.}
Experiences with vibe coding are broadly similar across experience levels. Both positive 
experiences and challenges are reported at comparable levels regardless of prior 
programming experience. This contrasts with the motivational differences observed in RQ1, 
suggesting that once participants engage with vibe coding, the experience itself converges 
across experience levels.

\begin{table}[t]
\centering
\caption{Experience with vibe coding across experience groups.}
\label{tab:rq2_experience}
\begin{tabular}{lccccc}
\toprule
\textbf{Theme}
& \textbf{Non-dev} & \textbf{Novice} & \textbf{Pro}
& \textbf{$p_{BH}$} & \textbf{$\varepsilon^2$} \\
\midrule

Instant Success \& Flow
& 3 (1) & 3 (1) & 3 (2)
& .892 & .000 \\

Prompt Struggle \& Iteration
& 4 (2) & 4 (1.75) & 4 (1)
& .892 & .000 \\

Code Breakdown or Abandonment
& 4 (2.25) & 3 (2) & 3.5 (3)
& .892 & .000 \\

Fun \& Creative Satisfaction
& 4 (2) & 4 (2) & 4 (1)
& .437 & .020 \\

AI Hallucinations
& 4 (1) & 4 (2) & 4 (1)
& .437 & .012 \\

Confusion or Misunderstanding
& 4 (2) & 3 (1) & 4 (2)
& .892 & .000 \\

\bottomrule
\end{tabular}
\end{table}

\subsection{Perceived Code Quality of the AI-Generated Code (RQ3)}
\begin{figure}[t]
    \centering
    \includegraphics[width=\linewidth]{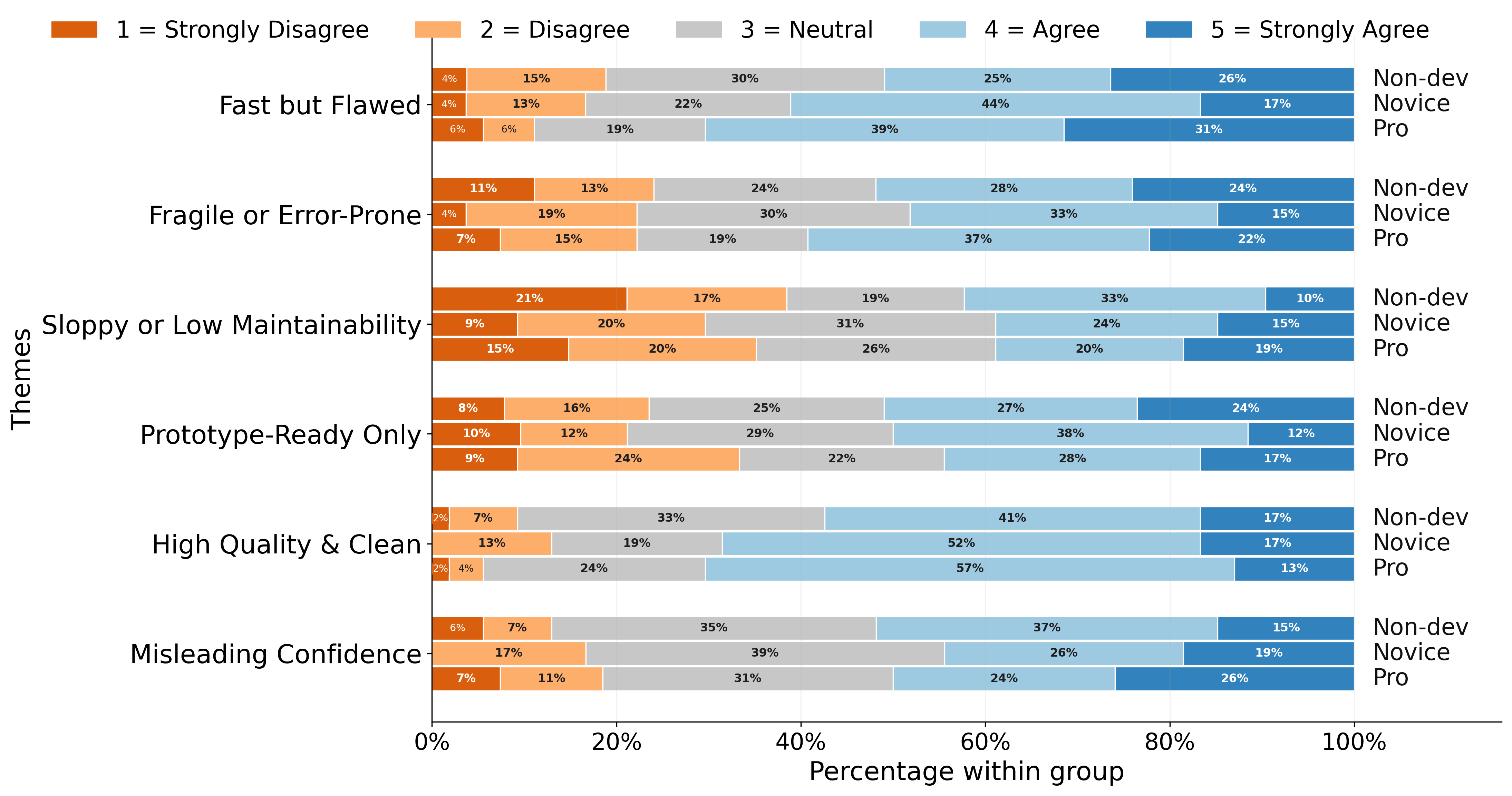}
    \caption{Distribution of perceived code quality responses across groups.}
    \label{fig:rq3_quality}
\end{figure}

RQ3 examined participants' perceptions of the quality, robustness, and production 
readiness of \aigc{} produced through vibe coding. Figure~\ref{fig:rq3_quality} shows 
the distribution of responses across experience groups, while Table~\ref{tab:rq3_quality} 
summarises medians and interquartile ranges. No statistically significant differences 
were observed across experience levels for any theme ($p_{\text{BH}} = .884$ for all, 
$\varepsilon^2 \leq .008$), and responses were broadly consistent across groups, ranging 
between neutral and agreement ($\text{Mdn} = 3$--$4$). Participants held a cautiously mixed view of \aigc{} quality. On the positive side, participants moderately agreed that \aigc{} can sometimes be well-structured and clean. At the same time, participants consistently acknowledged its limitations: \aigc{} is perceived as useful for rapid prototyping but often flawed, fragile or error-prone, and 
more suitable for demos than production deployment. Participants also indicated that 
\aigc{} may appear correct while concealing deeper issues, and reported neutral 
perceptions regarding its maintainability and structural quality.

\noindent\textbf{Overall Pattern.}
Perceived code quality is broadly similar across experience levels. Participants consistently recognise both the strengths and limitations of \aigc{}, indicating that these perceptions are largely independent of prior programming experience. Notably, a general awareness of quality issues is shared across all groups, even though the ability to evaluate and act on those issues appears to vary with experience, as reflected in the QA practice differences observed in RQ4.

\begin{table}[t]
\centering
\caption{Perceived code quality of vibe-coded outputs (RQ3) across experience groups.}
\label{tab:rq3_quality}
\begin{tabular}{lccccc}
\toprule
\textbf{Theme}
& \textbf{Non-dev} & \textbf{Novice} & \textbf{Pro}
& \textbf{$p_{BH}$} & \textbf{$\varepsilon^2$} \\
\midrule

Fast but Flawed
& 4 (2) & 4 (1) & 4 (2)
& .884 & .008 \\

Fragile or Error-Prone
& 4 (1) & 3 (1) & 4 (1)
& .884 & .000 \\

Sloppy or Low Maintainability
& 3 (2) & 3 (2) & 3 (2)
& .884 & .000 \\

Prototype-Ready Only
& 4 (1) & 3.5 (1) & 3 (2)
& .884 & .000 \\

High Quality \& Clean
& 4 (1) & 4 (1) & 4 (1)
& .884 & .000 \\

Misleading Confidence
& 4 (1) & 3 (1) & 3.5 (1.75)
& .884 & .000 \\

\bottomrule
\end{tabular}
\end{table}

\subsection{QA Practices when Vibe Coding (RQ4)}
\label{Result-RQ4}
\begin{figure}[t]
    \centering
    \includegraphics[width=\linewidth]{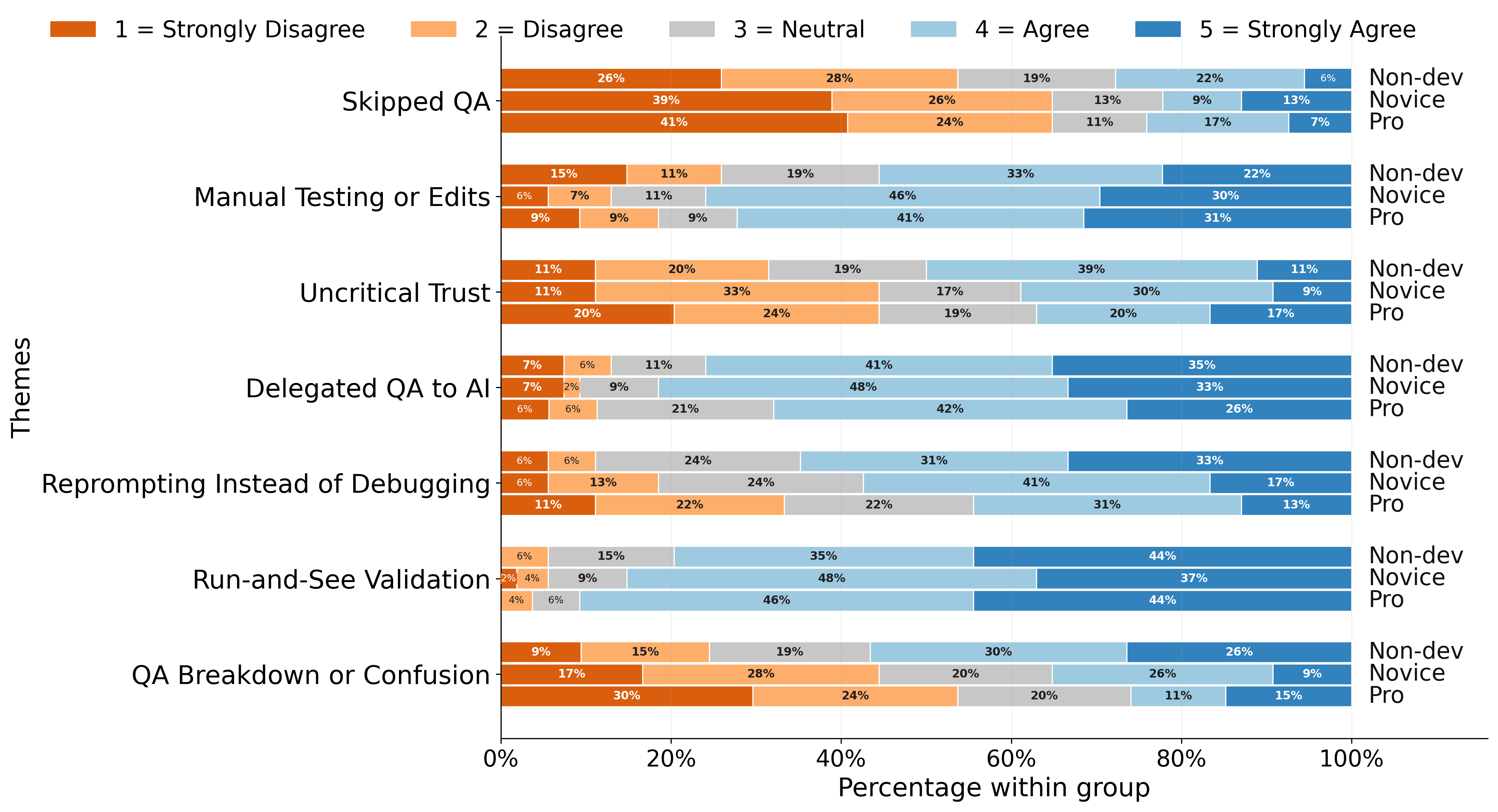}
    \caption{Distribution of QA practice responses across experience groups.}
    \label{fig:rq4_qa}
\end{figure}

RQ4 examined how participants validate, test, and troubleshoot \aigc{} during vibe coding. Figure~\ref{fig:rq4_qa} shows the distribution of responses across experience groups, while Table~\ref{tab:rq4_qa} summarises medians and interquartile ranges. In contrast to RQ2 and RQ3, statistically significant differences across experience levels were observed for selected QA practices. Two themes showed statistically significant differences: \textit{Reprompting Instead of Debugging} ($p_{\text{BH}}=.037$, $\varepsilon^2=.045$) and \textit{QA Breakdown or Confusion} ($p_{\text{BH}}=.012$, $\varepsilon^2=.068$). As shown in Table~\ref{tab:pairwise_all}, non-developers report higher reliance on reprompting rather than manually debugging errors (\textbf{\textit{Reprompting Instead of Debugging}}) than professionals ($p_{\text{adj}}=.008$), while novices did not differ significantly from either group. For \textbf{\textit{QA Breakdown or Confusion}} (capturing the extent to which participants give up on debugging because they cannot understand the \aigc{}), Non-developers report higher agreement than both novices ($p_{\text{adj}}=.027$) and professionals ($p_{\text{adj}}=.002$), with no significant difference between novices and professionals. All remaining QA practices (Skipped QA, Manual Testing or Edits, Uncritical Trust, Delegated QA to AI, and Run-and-See Validation) did not differ significantly across groups ($p_{\text{BH}} \geq .259$, $\varepsilon^2 \leq .015$), and manual testing and editing was commonly reported across all groups ($\text{Mdn} = 4$).

\begin{figure}[t]
    \centering
    \includegraphics[width=1\linewidth]{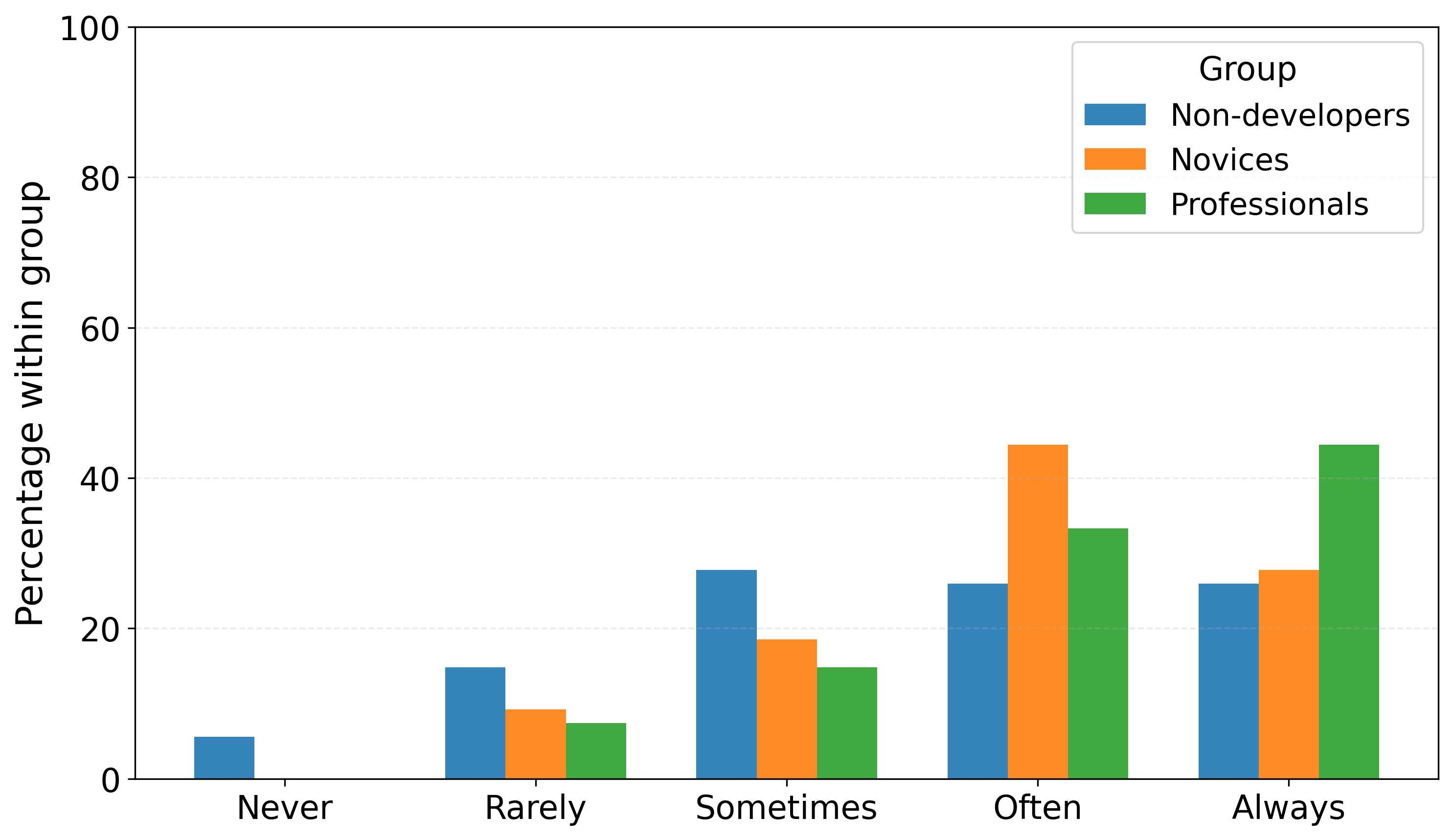}
   \caption{Frequency of checking AI-generated code across experience groups.}
    \label{fig:rq4_check}
\end{figure}

\noindent\textbf{Frequency of Checking AI-Generated Code.}
As shown in Figure~\ref{fig:rq4_check}, checking frequency varied across experience groups ($\chi^2(8)=16.04$, $p=.042$; Kruskal--Wallis $p=.013$). Professionals reported the highest rates of consistent checking, with approximately 45\% reporting that they always check \aigc{} before using it. Novices were most concentrated in the \textit{often} category ($\approx$45\%), while non-developers were more evenly distributed across lower-frequency categories and were the only group to report never checking \aigc{} prior to use.

\noindent\textcolor{black}{\textbf{ Regression Analysis.}
To examine whether the QA differences observed above are explained by experience-group membership itself or by the underlying dimensions of experience (usage), we ran three ordinal logistic regressions, one per significant QA outcome (see Section~\ref{Meth-QuantStatAnaly}). The three QA regression models (Table~\ref{tab:rq4_regression}) all satisfied the proportional-odds assumption (Brant test, all $p>.78$), with negligible multicollinearity (all GVIF $<1.3$). Across all three models, experience group did not significantly predict QA behaviour once the three usage dimensions of experience (weekly vibe coding hours, adoption duration, and weekly independent non-vibe coding activity) were included. Instead, the significant predictors were the usage dimensions themselves. Greater adoption duration and greater independent coding activity were each associated with less reprompting, and greater independent coding activity (coding without vibe coding) was also associated with more frequent checking of \aigc{}. This indicates that the QA differences are not arbitrary group effects but are shaped by accumulated coding practice and tool exposure, the specific components of experience that distinguish the three groups.}

\begin{table}[t]
\centering
\textcolor{black}{
\caption{Ordinal logistic regression of QA outcomes on experience group and usage variables. Cells report odds ratios (95\% CI). Non-developers are the reference group.}
\label{tab:rq4_regression}
\begin{tabular}{lccc}
\toprule
\textbf{Predictor} & \textbf{Reprompting} & \textbf{QA Breakdown} & \textbf{Checking Freq.} \\
\midrule
Novice (vs non-dev)        & 1.00 (0.65--1.54) & 0.75 (0.49--1.14) & 1.25 (0.81--1.93) \\
Professional (vs non-dev)  & 1.01 (0.61--1.68) & 0.76 (0.46--1.26) & 1.40 (0.83--2.34) \\
Weekly vibe hours          & 0.84 (0.69--1.03) & 0.85 (0.70--1.03) & 1.02 (0.84--1.25) \\
Adoption duration          & \textbf{0.84 (0.72--0.99)} & 0.89 (0.76--1.03) & 1.01 (0.86--1.18) \\
Non-vibe coding hours      & \textbf{0.85 (0.75--0.97)} & 0.92 (0.81--1.04) & \textbf{1.15 (1.01--1.31)} \\
\bottomrule
\multicolumn{4}{l}{\footnotesize Bold $=p<.05$.} \\
\end{tabular}
}
\end{table}

\noindent\textbf{Overall Pattern.}
QA practices show selective differences across experience levels. Differences emerge primarily in debugging-related behaviours, where less experienced participants report greater reliance on reprompting and higher levels of confusion when faced with \aigc{} they cannot understand. More experienced participants report lower levels on both measures and check \aigc{} more frequently and consistently. \textcolor{black}{The regression analysis further showed that these differences are driven by accumulated coding practice and tool exposure rather than by experience-group category alone. Together, these results indicate that while perceptions of code quality are broadly shared, the ability to evaluate and respond to quality issues remains dependent on experience, specifically on accumulated coding practice and tool exposure.}

\begin{table}[t]
\centering
\caption{QA practices during vibe coding (RQ4) across experience groups.}
\label{tab:rq4_qa}
\begin{tabular}{lccccc}
\toprule
\textbf{Theme}
& \textbf{Non-dev} & \textbf{Novice} & \textbf{Pro}
& \textbf{$p_{BH}$} & \textbf{$\varepsilon^2$} \\
\midrule

Skipped QA
& 2 (2.75) & 2 (2) & 2 (2)
& .485 & .000 \\

Manual Testing or Edits
& 4 (1.75) & 4 (1) & 4 (2)
& .259 & .015 \\

Uncritical Trust
& 3.5 (2) & 3 (2) & 3 (2)
& .485 & .000 \\

Delegated QA to AI
& 4 (1) & 4 (1) & 4 (2)
& .485 & .000 \\

\textbf{Reprompting Instead of Debugging}
& \textbf{4 (2)} & \textbf{4 (1)} & \textbf{3 (2)}
& $\mathbf{.037}$ & \textbf{.045} \\

Run-and-See Validation
& 4 (1) & 4 (1) & 4 (1)
& .636 & .000 \\

\textbf{QA Breakdown or Confusion}
& \textbf{4 (2)} & \textbf{3 (2)} & \textbf{2 (2.75)}
& $\mathbf{.012}$ & \textbf{.068} \\

\bottomrule
\end{tabular}
\end{table}

\subsection{Improving Trust in Code Generated by Vibe Coding}
\label{sec:rq5_q15}

\textcolor{black}{
To complement our analysis, we analysed responses to an open-ended question asking participants what features would make \aigc{} (vibe coding output) easier and safer to use or trust. We received responses from all \AnalysedSampleSize{} participants. We conducted a thematic analysis of all responses and grouped all responses into four recurring themes (responses could be assigned to multiple themes).}
\textcolor{black}{
The most prominent theme was \textit{reliability and correctness}. Participants emphasised the need for dependable outputs, early error detection, and functionally correct behaviour, requesting features such as ``clear explanations of what the code does, automatic error or bug warnings, and easy suggestions for improvements or fixes.'' A second theme, \textit{workflow support and control}, reflected a need for process-level visibility during development, specifically the ability to observe how generated code executes and to receive step-by-step guidance as the tool produces it, with one participant wanting tools that ``give reasons for the code as we go.'' A third theme, \textit{transparency and understandability}, focused on comprehension of the code artifact itself: participants wanted clearer documentation and explanations of what the generated code does and why specific implementation choices were made. The least common theme, \textit{safety, privacy, and risk control}, reflected concerns about overly complex or risky outputs, with participants preferring tools that ``do not suggest further and complicated output.'' These feature expectations were broadly consistent across the three experience groups, suggesting they are largely shared regardless of prior programming experience.}

\subsection{Summary of the Results}
The results show that experience level does not uniformly shape vibe coding practices; instead, its influence is concentrated in specific parts of the vibe coding process. Differences are most evident in how participants \emph{approach} and \emph{handle failures} when vibe coding. 
Non-developers rely more on AI-led code generation and view it as an enabling tool. Novices emphasise learning from the code generation and view it as an enabling tool for a better software development experience. Professionals engage with vibe coding more through interactive prompting and structured guidance, yet perceive the experience and quality of \aigc{} in ways broadly similar to those of less-experienced groups.
Across groups, participants recognise a similar balance of strengths (e.g., usefulness for rapid prototyping and well-structured output) and limitations (e.g., fragility, hidden bugs, and limited production readiness). The clearest differences emerge in quality assurance and debugging. Less experienced participants rely more on reprompting and report greater confusion, whereas more experienced participants report less reprompting, less debugging breakdown, and more frequent assessments of the \aigc{}. This suggests that experience level largely influences how participants respond to errors and limitations in \aigc{}, rather than how they perceive them.\textcolor{black}{The regression analysis showed that these QA differences are driven by accumulated coding practice and tool exposure rather than by experience-group category alone.}

The results also show varying expectations across the three groups, with participants consistently emphasising reliability, transparency, workflow support, and safety. 

\section{Discussion}
\label{discussion}

\begin{figure}
    \centering
    \includegraphics[width=1\linewidth]{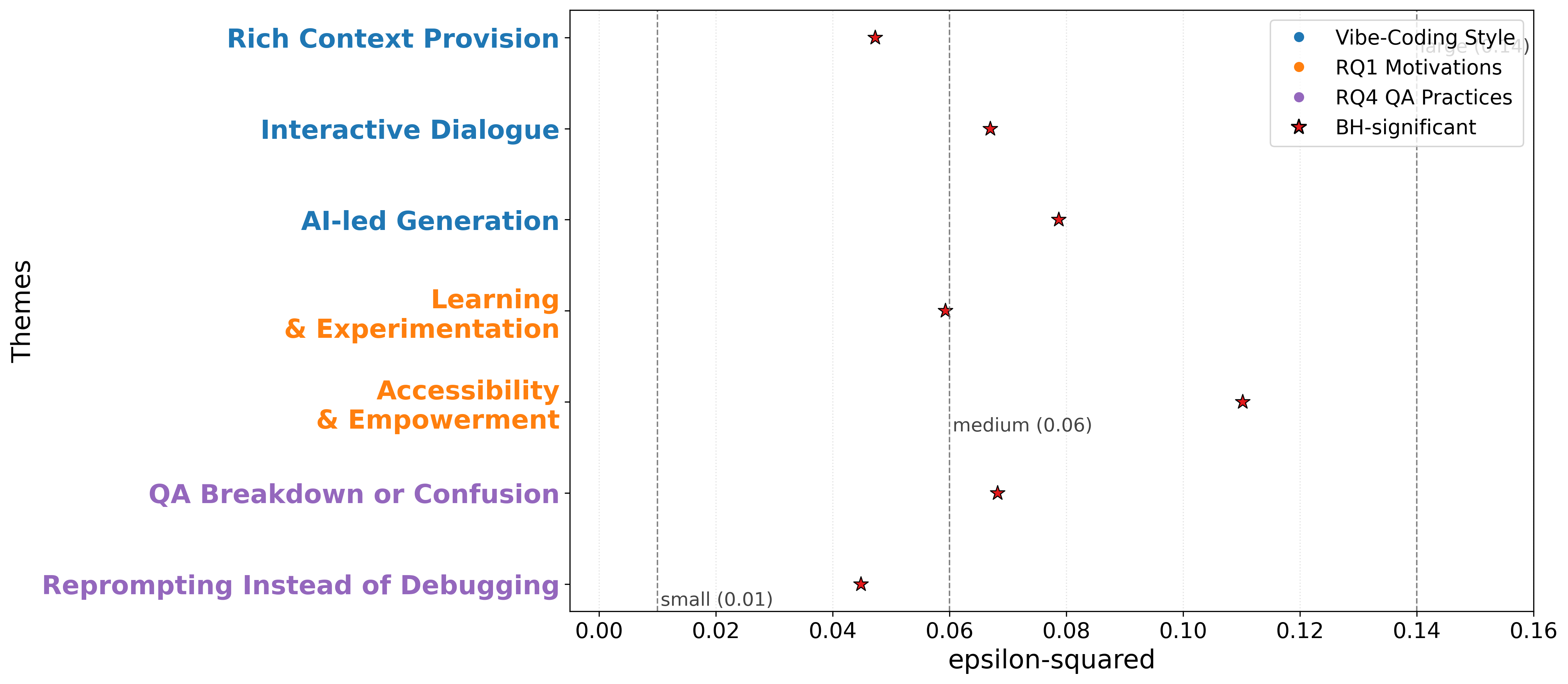}
    \caption{Experience Influences Behaviour but Not Perception in Vibe Coding: Effect Sizes Across RQ1--RQ4 and Interaction Style}
    \label{fig:EffectSizes}
\end{figure}

\subsection{Key Observations}

\noindent \textbf{Selective Influence of Experience:}
Across all research questions, a consistent pattern emerges: experience primarily influences behavioural themes, while reported experiences and perceived code quality remain largely stable. As shown in Figure~\ref{fig:EffectSizes}, effect sizes are negligible for \textit{experiences} (RQ2) and \textit{perceived code quality} (RQ3), but small-to-moderate for \textit{motivations} (RQ1), interaction style, and selected \textit{QA practices}  (RQ4). 

This pattern is also evident in vibe coding interaction styles, where experience shapes how users engage with \AItools{}, shifting from passive reliance on these tools, towards more iterative and context-rich interaction, while core practices such as planning and verification remain stable.
In turn, 
awareness of risks in \aigc{}, such as fragility, hidden defects and bugs, and misleading correctness, is broadly shared at a general level between all three user groups. However, the ability to respond to these risks varies with experience. \textcolor{black}{We interpret this pattern as a potential \emph{perception--action gap}: an apparent decoupling between recognising potential issues and effectively addressing them. Because our measures are based on self-report rather than observed behaviour, we present this as an interpretation consistent with our data and a hypothesis for future behavioural studies to test, rather than a directly validated finding (Section~\ref{ThreatstoValidity}).} This observation aligns with prior findings showing that developers often rely on \AItools{} despite recognising its limitations~\cite{vaithilingam2022expectation,wang2024investigating,yardim2026ai}.\\




\noindent \textbf{Vibe Coding as a Partially Democratising Practice:}
The absence of significant differences in RQ2 (\textit{experiences}) and RQ3 (\textit{perceived code quality}) is a notable result. Participants across all experience levels report similar interaction patterns, including flow, iterative prompting, and general awareness of code limitations. This convergence suggests that participants across all experience levels report broadly similar experiences when vibe coding, in which flow, iterative reprompting, awareness of hallucinations, and creative satisfaction emerge at comparable levels regardless of prior programming background, consistent with the goals of end-user programming to lower barriers to software creation \cite{ko2011state,scaffidi2005estimating}.

However, this convergence operates primarily at the level of perception rather than action. While access to code generation and high-level awareness of risks appears broadly distributed, the ability to evaluate, debug, and validate outputs remains uneven across experience levels. This indicates a form of \emph{partial democratisation}, i.e., rather than eliminating barriers, vibe coding shifts them from code production to other barriers such as decomposing problems, specifying intent clearly, and valuating the correctness and robustness of \aigc{}. 

Alternative explanations should also be considered. Similar perceptions of code quality may reflect shared exposure to discourse about AI limitations, rather than equivalent evaluative capability.
\textcolor{black}{Relatedly, our items measured awareness of code limitations at a general level (e.g., whether \aigc{} is error-prone or fragile) rather than probing specific risk categories such as security vulnerabilities. Similar responses to these high-level items therefore indicate broadly shared general awareness, but not necessarily an equivalent depth or accuracy of risk understanding across experience groups. }

Likewise, behaviours such as reprompting may represent an efficient interaction strategy rather than a deficit. Nevertheless, the observed divergence in QA practices suggests that, even when risks are recognised, the capacity to act on this awareness remains experience-dependent. \\

\noindent \textbf{Divergent Entry Points: Motivation and Interaction Style:}
Differences in specific motivations and interaction styles reflect distinct modes of engagement shaped by experience. Non-developers are primarily driven by accessibility and tend to rely on AI-code generation with minimal structure. Novices emphasise learning and experimentation, using vibe coding as an exploratory environment. In contrast, professional developers are more goal-oriented, engaging in iterative dialogue and providing richer contextual input.

These differences suggest that vibe coding is not a uniform practice but a flexible interaction paradigm that adapts to coder intent and prior experience. Our findings regarding vibe coding application context  (Section~\ref{Res-VCappContex}) further support this interpretation. Non-developers reported the highest levels of hobby-oriented use, suggesting that vibe coding serves as an accessible entry point for creative exploration for this group. In contrast, our survey results showed that novices engage more in goal-directed contexts such as learning and assignments, whereas experienced developers predominantly use it for professional tasks. This pattern indicates that differences in experience are reflected not only in behaviour, but also in how and why vibe coding is adopted.\\

\noindent \textbf{Divergence in Quality Assurance Behaviour:}
The perception--action gap is most evident in QA practices. Less-experienced participants rely more on reprompting and report greater difficulty debugging, whereas experienced developers more consistently verify and evaluate outputs. This divergence occurs despite similar perceptions of code quality, reinforcing that recognising potential issues does not necessarily translate into effective corrective action. \textcolor{black}{A follow-up regression analysis reinforced this view: once accumulated coding practice and tool exposure were modelled directly, experience-group membership no longer carried independent predictive power. This indicates that QA behaviour is shaped by the specific dimensions of experience (coding practice and exposure) rather than by group category alone.}

This pattern is consistent with prior work showing that debugging and code comprehension depend on experience, mental models, and the use of debugging and comprehension strategies 
\cite{baltes2018towards,parnin2011automated,siegmund2014understanding,latoza2010developers}. In the context of AI-assisted development, these capabilities remain critical but are no longer directly reflected in perceived code quality.

\subsection{Implications}
We discuss below several implications for vibe coding in research and practice.  \\

\noindent \textbf{Decoupling Perception and Action in Software Engineering Expertise:}
Prior work on expertise suggests that the ability to recognise and resolve problems develops together through experience and practice \cite{ericsson1993role}. Our findings refine this view in the context of AI-assisted development. While general awareness of potential issues appears broadly accessible across experience levels, the ability to evaluate, debug, and verify code remains strongly experience-dependent. This indicates that AI-assisted development reshapes how expertise is expressed: perceptual awareness becomes more widely distributed, whereas action-oriented capabilities continue to differentiate levels of expertise.

Future research can distinguish between awareness-level and resolution-level competences when studying expertise in AI-assisted contexts, and examine how tools and software training can better support the transition between them.\\

\noindent \textbf{Implications for AI-Coding Tools:}
The perception--action gap highlights a limitation of current \AItools{}, which primarily support code generation but provide limited support for evaluation and verification. Addressing this gap requires those tools to actively support users in translating awareness into effective action, particularly for less experienced coders.

Building on these findings, we derive three design directions from both the observed QA behaviours (RQ4) and participants’ reported needs for tool support. 
First, \textit{proactive error surfacing}, where tools identify and communicate potential issues at generation time in ways that guide concrete corrective steps. Second, \textit{explanatory support}, providing context-sensitive explanations that help coders understand not only what the code does, but how to evaluate and adapt it. Third, \textit{uncertainty signalling}, explicitly communicating assumptions and limitations to prompt verification behaviours rather than passive acceptance.
These capabilities should be adaptive to coder experience, providing stronger scaffolding for less experienced coders while remaining unobtrusive for experts, in line with prior work on scaffolding and intelligent tutoring systems \cite{reiser2018scaffolding,weintrop2019transitioning,vanlehn2011relative,winkler2018unleashing}.\\

\noindent \textbf{Implications for Practice and Education:}
The perception--action gap suggests that adopting vibe coding without structured QA processes introduces risk, particularly in mixed-experience teams \cite{perry2023users,vaithilingam2022expectation}. Even when coders recognise limitations in \aigc{}, they may lack the skills required to act on this awareness effectively. Establishing explicit verification practices, including validation steps and human code review, is therefore essential \cite{bacchelli2013expectations}.

For education, these findings highlight the need to treat evaluation of \aigc{} as a distinct skill. Prior work has shown that \aigc{} can contain security vulnerabilities and functional defects, including subtle bugs that may not be immediately visible during execution \cite{fu2025security,ambati2024navigating,pearce2025asleep,gao2025survey,tambon2025bugs}. Instruction should extend beyond prompting to include testing outputs, identifying edge cases, challenging assumptions, and reasoning about code behaviour. Developing these competencies is critical for enabling effective use of AI-assisted development tools.

%
%
%
%
%
%

\section{Threats to Validity}
\label{ThreatstoValidity}
We note several potential validity threats that could affect the interpretation of our findings. We list each threat in the following:

\smallskip
\noindent \textbf{Self-reported experience classification.}
Participants were assigned to experience groups based on a self-classification question, introducing a risk of misclassification if participants overestimate or underestimate their own background. To reduce this risk the response options did not ask participants to subjectively rate their ability; instead, each option described a recognisable scenario anchored to concrete activities (e.g., the non-developer option specified ``no formal programming training or professional experience, but having used AI tools to generate or modify code''; the professional option specified ``professional or academic experience writing, maintaining, or reviewing code as part of a job, research, or formal development work''). This formulation allowed participants to map themselves onto the option closest to their actual experience rather than their self-perceived skill level. Furthermore, the experience self-classification question was administered twice in separate sessions approximately four weeks apart, allowing us to verify the stability of group assignments (Section~\ref{Meth-Data-collection}); the matched-sample agreement of 74.0\%, with the majority of changes drifting toward the middle ``novice'' category and only one participant moving between the two most distinct categories, indicates that misclassification at the boundaries of the most distinct groups is unlikely to drive the observed group differences.

\smallskip
\noindent \textbf{Construct coverage and operationalisation.}
Our study constructs (motivations, experiences, perceived code quality, and QA practices) were derived from themes identified in our previous work~\cite{fawzy2025vibe}, grounding the survey in real-world practice. While this supports content validity, some aspects of vibe coding may remain underrepresented. To check coverage, we included an optional open-ended response for each construct section and analysed the responses to identify any constructs not captured by our items; no additional dimensions emerged (Section \ref{Meth-ConstructCoverageVer}).

\smallskip
\noindent \textbf{Common method bias.}
All measures rely on participants' self-reports collected via a single survey at a single time point, increasing the risk of self-reporting and common method biases~\cite {podsakoff2003common}. Participants may overstate the extent of careful QA practices or understate reliance on \aigc{}, and consistent response tendencies may inflate the observed relationships between constructs. To reduce these effects, we (i) collected responses anonymously to lower social desirability pressure, (ii) separated the experience-classification stage from the main survey, (iii) organised constructs into distinct sections to reduce carry-over effects, (iv) used multi-item Likert scales per construct rather than single items, allowing patterns to be assessed at the construct level rather than relying on any one response, and (v) provided a consistent definition of vibe coding at the start of the survey to ensure shared understanding among participants. Where possible, items were phrased in behavioural rather than evaluative terms.

\smallskip
\noindent \textbf{Perception--action gap interpretation.}
The perception--action gap is inferred from differences between reported awareness of \aigc{} limitations and reported QA behaviours. A key threat is that this pattern reflects self-reporting differences rather than actual behaviour (e.g., professionals may have underreported weaker practices, or less-experienced participants may have reported them more explicitly). To strengthen the interpretation, we examined the consistency of this pattern across multiple independent constructs (RQ2, RQ3, and RQ4) and across interaction styles. The recurrence of the pattern across different measures suggests that it is not an artefact of any single item or construct. \textcolor{black}{Future work should validate this interpretation by examining actual behaviour, for example, through repository analysis or observational studies of how vibe coders evaluate and verify \aigc{}}.

\smallskip
\noindent \textbf{Sample composition and generalisability.}
Participants were recruited via Prolific, an online crowdsourcing platform. Although Prolific samples are more diverse than typical student samples~\cite{peer2017beyond}, they may overrepresent individuals familiar with online platforms and survey-based tasks. The sample also reflects international participation with varied geographic and educational backgrounds, which may introduce cultural variation in attitudes towards \AItools{} and in self-assessment. The inclusion of three distinct experience groups broadens coverage, but generalisation beyond similar populations should be made with caution.

\smallskip
\noindent \textbf{Statistical robustness.}
The analysis involves multiple comparisons within each research question, increasing the risk of false positives. We applied the Benjamini--Hochberg procedure to control the false discovery rate, and we interpret results at the construct level, emphasising consistent patterns across related items and their effect sizes rather than isolated significant findings. We additionally excluded implausibly rapid responses using a completion-time threshold; a sensitivity analysis with a stricter threshold produced consistent patterns, supporting the robustness of the findings.

\section{Conclusion}
\label{conclusion}
This paper presents a survey of vibe coding practices, comparing responses from \AnalysedSampleSize{} participants, representing three vibe coding user groups: non-developers, novice developers, and professional developers. We investigated vibe coding practices among those groups across four dimensions: motivations, experiences, perceived code quality, and QA practices followed. 

Our analysis of the survey samples shows that prior programming experience shapes vibe coding selectively rather than uniformly. Reported experiences and perceptions of \aigc{} quality were largely consistent across groups, indicating a broadly shared general awareness of both the benefits and the limitations of vibe coding. Group differences instead concentrated on three areas: \emph{why} participants engage in vibe coding, \emph{how} they interact with \AItools{}, and \emph{how} they respond to issues in the code. Non-developers approached vibe coding as an accessibility-enabling AI-led practice. Novices treated it as an exploratory environment for learning. professionals adopted a more goal-oriented, iterative, and context-rich style, with greater use in work-related contexts. The clearest divergence appeared in QA practices, where less-experienced participants more often replaced debugging with reprompting using the same tools, and reported more frequent breakdowns when \aigc{} was difficult to understand, whereas professionals verified outputs more consistently.
 
We interpret these observations as a potential \emph{perception--action gap}. A general awareness of risks in \aigc{} appears broadly distributed across experience levels, but the capacity to act on that awareness through evaluation, debugging, and verification appears to remain experience-dependent. This reframes vibe coding as a \emph{partially democratising} practice. Access to code production and high-level awareness of risks expands across user groups, yet evaluation and verification continue to differentiate experience levels. The risks of vibe coding, therefore, depend not only on the \AItools{}, but also on the user's capability to act on what is already perceived. More broadly, this points to a partial decoupling between perceptual awareness and action-oriented competence in AI-assisted development.
 
 
Future work should examine the perception--action gap through behavioural studies that move beyond self-report, investigate how scaffolding mechanisms in \AItools{} influence verification behaviour across experience levels, and evaluate educational interventions targeting the assessment of \aigc{}. As vibe coding continues to broaden participation in software creation, ensuring that the ability to evaluate what is produced keeps pace with the ability to produce it will be central to its safe and effective adoption.



\bibliographystyle{ACM-Reference-Format} 
\bibliography{References}



\end{document}